\documentclass[aps,preprint,a4paper,showpacs,amsmath,amssymb,floatfix,nofootinbib,superscriptaddress]{revtex4}  

\usepackage{graphicx}
\usepackage{amsmath}
\providecommand{\e}[1]{\ensuremath{\times 10^{#1}}}

\begin{document}

\title{Molecular Dynamics Simulations of Weak Detonations}
\author{Morag Am-Shallem}
\affiliation{The Fritz Haber Research Center, The Hebrew University, Jerusalem 91904, Israel}
\author{Yehuda Zeiri}
\affiliation{Department of Biomedical Engineering, Ben-Gurion University, 
Beer-Sheva 84105, Israel}
\affiliation{Chemistry Department, NRCN, P.O. Box 9001, Beer-Sheva 84190, Israel}
\author{Sergey V. Zybin}
\affiliation{Materials \& Process Simulation Center, California Institute of Technology, 
Pasadena, California 91125}
\author{Ronnie Kosloff}
\affiliation{The Fritz Haber Research Center, The Hebrew University, Jerusalem 91904, Israel}

\begin{abstract}
Detonation of a three-dimensional reactive non-isotropic molecular crystal is modeled
using molecular dynamics simulations.
The detonation process is initiated by an impulse, followed by the creation of a stable fast reactive shock wave.
The terminal shock velocity is independent of the initiation conditions.
Further analysis shows supersonic propagation decoupled from the dynamics of the decomposed material
left behind the shock front. The dependence of the shock velocity 
on crystal nonlinear compressibility resembles solitary behavior.
These  properties  categorize the phenomena as a weak detonation. 
The dependence of the detonation wave on microscopic potential parameters 
was investigated. 
An increase in detonation velocity with  the reaction exothermicity reaching a saturation value is observed. 
In all other respects the model crystal exhibits 
typical properties of a molecular crystal.
\end{abstract}


\maketitle

\section{Introduction}

Explosives are characterized by a detonation wave which propagates
through the material. After initiation, the
velocity of the detonation front reaches a steady state that
exceeds the speed of sound in the material. The present paper is
devoted to the analysis of a model solid explosive with the purpose of
correlating the microscopic structure and interatomic forces to the bulk detonation
properties. The investigation is based on a classical molecular dynamics
simulation with a simple force field. The initial goal was to
construct a first principle model which is able to qualitatively reproduce a
stable detonation wave. 
The microscopic parameters considered are the crystal structure, 
the intermolecular forces that stabilize this structure, 
and the intramolecular potentials which yield the driving chemical reaction. 
The present investigation unravelled a new type of a solitary-like detonation wave
which is directly driven by the one-step exothermic chemical decomposition. 
From a hydrodynamical perspective it can be classified as a weak detonation.

A common theoretical framework for simulating detonations relies on a continuum picture of 
the material properties using a hydrodynamic description,
and is based on the conservation of mass, momentum and energy
\cite{fickett2000detonation, zukas2002explosive, dremin1999toward, benartzi2003generalized}. 

In the Chapman-Jouguet (CJ) model, the detonation wave is considered 
as a discontinuity between "unburnt" and "burnt" material, 
assuming that the transition from one state to another does occur instantaneously. 
It assumes an infinite reaction rate and requires only the knowledge 
of the thermodynamic equilibrium state of the burnt material
to satisfy the conservation relations on the Hugoniot curve of compressed 
detonation products.
The resulting detonation wave has zero-width reaction zone and should satisfy 
the minimum entropy solution of the Rankine-Hugoniot equations which corresponds 
to {\em sonic condition} for downstream flow (downstream velocity with respect 
to the front, which equals the isentropic sound speed in reacted material).
Freely propagating one-dimensional detonation wave without the piston 
({\em unsupported detonation}) attains the steady-state velocity satisfying 
the sonic condition.
The CJ model is used extensively as an engineering tool enabling quick calculation
of the detonation velocity and pressure in the design of explosive devices. 

An extension of the CJ model was proposed by Zel\'dovich, von Neumann, and D\"oring
\cite{zeld1940,Neumann1942,doer1943} (ZND) incorporates finite-rate reaction kinetics 
resulting in a chemical reaction zone of finite length. 
It assumes that the unreacted explosive material behind the front is initially compressed
to a high pressure resulting in 
so-called von Neumann pressure spike.
Then the chemical reactions are initiated behind the spike, 
leading to the exothermic energy release, temperature increase and expansion of
the reaction products to the lower pressure. The structure and length of 
the reaction zone is determined by the equation of state of intermediate products.
After completion of the reactions, the products reach the final equilibrium state, 
at the end of the reaction zone. The final state 
determines the detonation velocity and depends only on the equation of state 
of the reaction products. For an unsupported detonation wave, this state satisfies 
the sonic condition as in the CJ model, and is called the CJ state.

In the ZND model, besides the minimum-entropy (or, correspondingly, minimum-velocity) solution to 
the conservation equations, there are two other possible detonation propagation regimes.
In a case of piston-supported detonation, the downstream flow can be subsonic 
in respect to the front and is called {\em strong or overdriven detonation}.
Another type of possible hydrodynamical solutions is termed weak detonation.
It is characterized by a downstream flow behind the detonation front which 
is supersonic in respect to the burnt material and effectively {\em decouples} from it 
(i.e. no disturbances from the burnt material can overtake the front). 
In addition, the weak detonation wave is characterized by only a moderate 
increase in density and pressure which is smaller than the CJ pressure 
on the detonation products Hugoniot.

Zel\'dovich \cite{zeld1942distribution} stresses that the weak detonation solution 
satisfies the boundary conditions of shock propagation and therefore is an admissible
hydrodynamical solution. He argues that the reason that the weak detonation phenomena
has not been observed experimentally is the unattainability of the steady state conditions
of this solution. 
In particular, a typical ZND route from the unreacted shock-compressed state 
to the burnt material gets trapped in the stable CJ point of minimum entropy production.
Von Neumann, however, has shown that if the Hugoniot curves of partially 
reacted products intersect one another then the weak detonation solution is possible \cite{Neumann1942}.
For example, explosives that have very rapid initial exothermic decomposition
followed by a slower endothermic reaction may exhibit such behavior sometimes called 
{\em pathological detonation}.
It has been also speculated by Tarver, based on a hydrodynamical equations, that in 
porous materials weak detonation could be stable \cite{Tarver1982chemicalEnergyRelease}.
Though standard ZND model assumes that chemical reactions are triggered by the high 
temperature due to the initial shock compression, it is also possible to consider 
an alternate ignition mechanism without preliminary shock heating at the von Neumann spike.
Zeldovich \cite{zeld1960} discussed possible initiation of weak detonation by 
triggering chemical reactions with a sequence of sparks artificially fired along a tube.

Can a first-principle molecular dynamics (MD) simulation converge to the hydrodynamical detonation model?

There has been a continuous effort to develop MD simulation methods of explosives 
\cite{rice1998msd, Sorescu2003mdem}. 
An interaction potential, known as the Reactive Empirical Bond Order potential (REBO),
was introduced in Ref. \cite{Brenner1993rebo}.
This potential has a complex functional form and several parameters.
Molecular dynamics simulations of detonation through a model 2D crystal using this potential,
showed agreement with the predictions of the CJ model \cite{rice1996rebohydro}.

A significant step was the development of a reactive force field which accounts for breaking
and forming chemical bonds during the passage of the shock wave through the solid \cite{van2001reaxff}.
Simulations of actual explosives, such as RDX, PETN and TATP, have been attempted 
\cite{ Strachan2003RDX, Strachan2005RDX, Dubnikova2005TATP, vanduin2005TATP,
Liu2009SiPETN, landerville2009PETN, zybin2009pentarythritol,zybin2010PETNsens}. 
The main goal is to compare the simulation output to known
experimental characteristics of the system, such as detonation velocity
and final reaction products distributions. These studies were devoted to
establish a realistic simulation scheme which converges to the 
framework of the hydrodynamic models.
It was found that the goal of reaching steady state detonation conditions requires a major computational effort. 
The phenomenon stabilizes only in a mezoscopic scale (micrometers) which requires very large scale calculations including many millions of atoms.
To establish such a method as a predictive tool, many such simulations should be performed and compared with experiment. 
A second round of improvements should then be applied to the force field. 
At present, this task is still in its infancy.

Schemes to bridge the gap between the hydrodynamical description of detonation and the MD approach have been explored. 
The idea is to replace a  group of molecules by a single mesoparticle with an 
internal thermodynamic degree of freedom \cite{stoltz2006reduced,maillet2006reduced}
or to describe a hydrodynamical cell by fictitious particles \cite{reed2006}.
In both these schemes individual molecular properties are overlooked.

In the present study a different MD approach was utilized. 
We limited the objective to establish a relationship between the forces governing the 
dynamics in the microscopic system and the macroscopic phenomena. 
The simple molecular force field employed is constructed to have only a small 
number of adjustable parameters. 
In a three dimensional model, we observed stable detonation waves, 
their characteristics being independent of the initial conditions. 
It required a more thorough investigation to identify that these detonation waves 
are of a different character from those in the standard ZND model.

A first indication of a new phenomena can be inferred from the one-dimensional (1-D) 
model of Peyrard et al. \cite{peyrard1985mmc}, constructed from a chain of 
unstable diatomic molecules. 
The dissociation reaction generated an accelerating detonation wave. 
In order to obtain a stable detonation wave artificial dissipation channels were added. 
Our analysis of their model revealed that a detonation wave in the 1-D chain
resembles a solitary wave. Further understanding requires a full 3-D 
crystal model which should supply a natural dissipative mechanism.

The initiation of detonation wave in solids is closely related to the propagation 
of a shock wave through a crystal lattice. Both waves are quasi-one-dimensional. 
Early MD simulations of a planar shock wave in a perfect fcc crystal
\cite{zybin1999LJcrystals, zybin2004}
revealed the formation of a solitary-like train at the shock front 
over a wide range of shock velocities. It has oscillatory structure 
and exhibits significant deviation from the thermal equilibrium inside the train.
In addition, the MD simulations of shocks in a perfect bcc crystal have shown
\cite{roth2005sol} 
that an isolated solitary wave can be emitted from the shock front and run ahead it
at significantly faster speed than the shock front velocity.
This links the present study to the subject of
solitary waves propagation in a discrete lattice. Solitons are characterized by 
a group velocity which is proportional to the amplitude of the wave. 
For a one-dimensional lattice we should refer to the work of Toda who established 
a relationship between the microscopic parameter of the repulsive part of
the potential and the group velocity \cite{Toda1976, Toda1983}. 
Holian has performed MD simulations of shock waves in the 1-D Toda 
lattice to study the dependence of solitary train structure at the shock front 
on the interaction potential parameters \cite{holian1981}.

Can the energy release reaction occur directly at the shock front as in the CJ model?
The prerequisite is a metastable molecule whose one-step exothermic decomposition 
is triggered by the shock front and its energy is immediately fed back 
into the detonation wave. 
By definition, shock wave propagation is faster than any linear elastic wave
in the same material. 
As a result it will always confront \textit{fresh} unperturbed molecules. 
For these molecules to contribute to the detonation event, their initial 
decomposition timescale should be similar or faster than the timescale 
determined by the front propagation velocity. 
Slower processes can take place behind the shock front. These processes 
may include thermal equilibration and heat dissipation as well as slower 
chemical reactions which lead to the final product distribution. 
If the slower reactions are endothermic, the weak detonation wave can be initiated.

In the present study we identified a diatomic molecule where the exothermic
decomposition is sufficiently fast and can proceed directly at the detonation front.
We observed a propagation of fast supersonic reactive wave initiated by the impulse 
impact in a three-dimensional perfect molecular crystal. We analyse the dependence 
of the detonation wave parameters on the interaction potential and the molecule 
decomposition kinetics. 
We find that the downstream flow is supersonic, and the density increase behind the front 
is very small, which is typical for a weak detonation. 
We also observe the decoupling of supersonic front propagation from the 
dissipative dynamics of the decomposed material left behind the front.
Our analysis suggests that the weak detonation wave in the simulated crystal
resembles behavior of a solitary wave propagating at the supersonic speed 
through the lattice.
The molecule decomposition is triggered at the detonation front where 
a significant thermal non-equilibrium exists. The released energy is directly 
channelled into acceleration of the atoms from the dissociated molecules
pushing solitary detonation front forward. Remarkably, the front propagation 
dynamics significantly depends on the relative orientation of the light and 
heavy atoms in the molecule with respect to the shock direction. 

The simulation results show a possibility of fast initiation of the molecule
decomposition reactions at the supersonic solitary wave front in contrast 
to the standard ZND model where the reactions take place behind the shock front, 
initiated by the shock compression.
Summarizing, the solitary front ignition can provide a mechanism for 
a direct transition to the weak detonation regime from the initial uncompressed state 
without going first to the high pressure state by the shock compression.

\section{The Computational Model}

An explosive is defined as a molecular crystal that can 
decompose to smaller particles, generating a stable detonation wave. 
The front of the detonation wave moves faster
than the speed of the linear compression wave in the molecular
crystal. Molecular dynamics simulation  of detonations requires a scheme of molecular forces, initial
geometry, boundary conditions and atomic masses. The model 
should be able to sustain a stable crystal structure.
Simulation of the system is based on the solution of the classical equations
of motion with sufficient accuracy, using a modified version of the MD code MOLDY,
which is described in appendix \ref{sec:moldy}.

\subsection{The Reactive Molecule Model}
\begin{figure}[htbp]  
\centering \includegraphics*[width=3.0in]{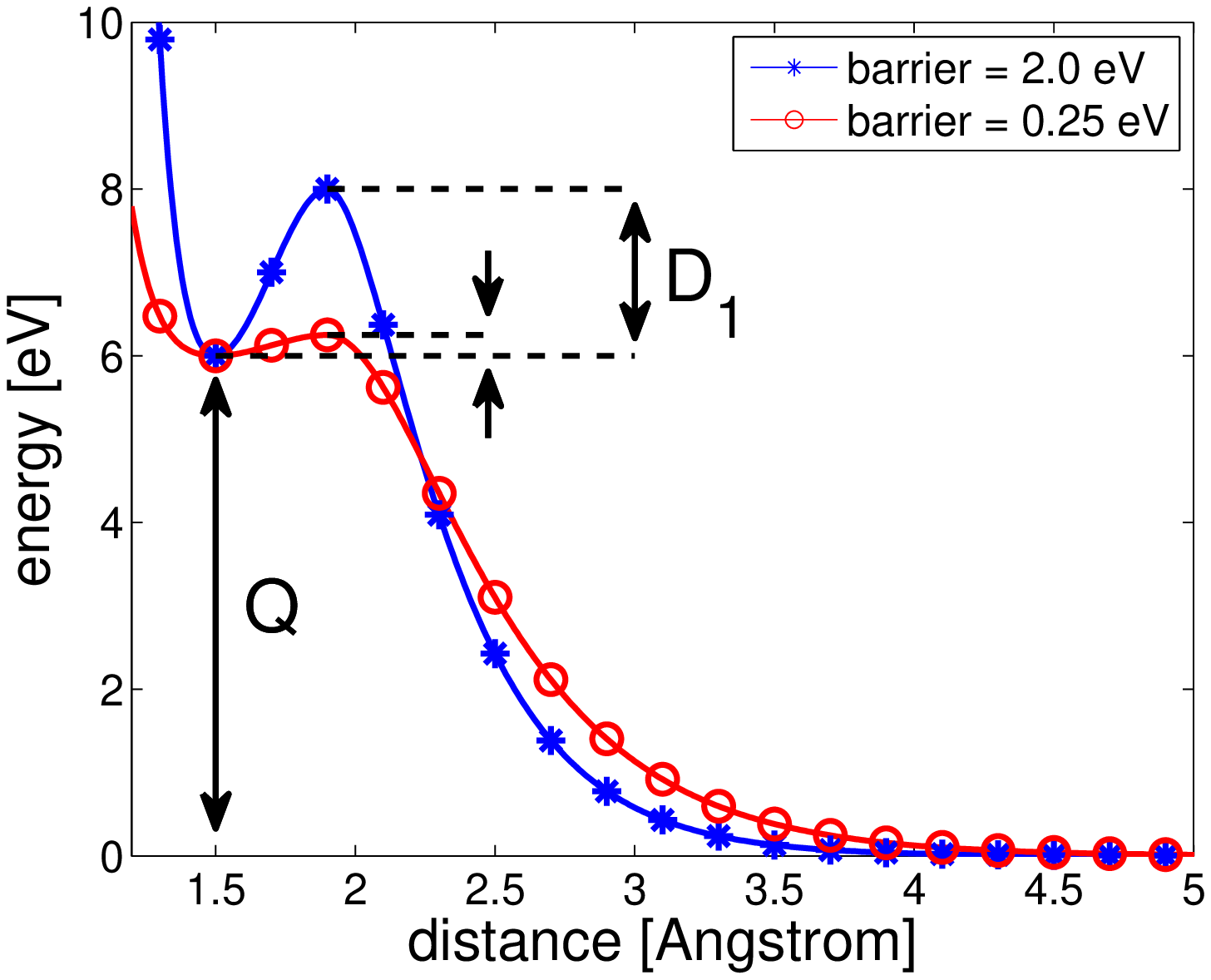}
 \caption{The reactive internal potential as a function of the  N-C distance. 
for two different ratios between the barrier height and the exothermicity. 
When the distance N-C is smaller than the barrier position 
(here at 1.9 \AA\ - the local maximum) the molecule is bound. 
When the distance is increased beyond this position the molecule dissociates. 
Initially, the N-C distance is at the equilibrium distance 
(here at 1.5 \AA\ - the local minimum).
$Q$ represents the exothermicity, 
and $D_1$ the barrier height for the N-C dissociation.
}
 \label{fig:exo_poten}
\end{figure}

The model crystal is represented by a slab of diatomic molecules arranged in a crystal structure. 
Each of the atoms can represent also a group of atoms.
The two effective particles will be marked as N and C, 
and the masses are 47 amu for N and 15 for C.
These notations and masses originated from nitromethane:
the N corresponds to the $\bf NO_2$ group, and the C to the $\bf CH_3$ group. 
The difference between the masses is essential in such reactive molecular model \cite{peyrard1985mmc}.
However, no other similarity to nitromethane exists in our model.
The chemical bond between N and C is designed to be metastable. 
When energized it can dissociate exothermally. 
The shape of the potential energy curve as a function of 
the N-C distance, is shown in figure \ref{fig:exo_poten}. 
When the distance N-C is smaller than the barrier
position the molecule is bound. However, when this distance is
increased beyond this position the molecule disintegrates. 
In an energetic perspective, the molecule absorbs sufficient energy such that its 
internal energy is higher than the energy barrier. 
Separation of the N-C atoms results in decomposition of the
molecule to its constituents accompanied by energy release. 
The evaluation of the potential and force during the trajectory calculations
was carried out by generating and storing a table of the potential and force values 
at different N-C distances, according to the potential shown in figure \ref{fig:exo_poten}, or similar potentials. 
A cubic spline interpolation was used to extract the values. 
There are some functional forms that can be used to generate such an exothermic reactive potential,
some of them are described in Ref. \cite{rice1998msd}.
We used a piecewise defined function to generate the potential, 
its functional forms are described in appendix \ref{exo_poten_func_form}. 

\subsection{The Molecular Crystal}

The molecular crystal was constructed by a lattice of N-C molecules with a bonding 
interaction between similar groups in neighboring molecules. 
The N entity interacts with other N and C interacts with other C entities. 
A Morse potential was chosen to describe these interactions:
\begin{equation}
V(r) = D\left(e^{-2\beta(r-r_0)} - 2e^{-\beta(r-r_0)} \right)
\label{eq:morse}
\end{equation}
The parameters used are shown in table \ref{tab:poten_param}. 
The initial distances between nearest neighbor molecules was chosen as $r_0=7.5 \text{\AA}$.
A long and narrow slab of this molecular crystal was assembled with FCC symmetry. 
Other structures studied, the BCC and simple cubic crystals, 
were found to be unstable with this type of pair potentials. 
This finding is consistent with analysis of stability of
Lennard-Jones crystals  \cite{dobbs1957tap}.

\begin{table}
 \begin{tabular}{|c|c|c|c|}
	\hline
  interaction & D  [eV] & $ \beta\ [\text{\AA}^{-1}] $ & $ r_0\ [\text{\AA}] $
  \\
	\hline
   N-N                  & 0.08  & 1.0  & 7.5 \\
   C-C                  & 0.016 & 1.0  & 7.5 \\
   N-C (inner part):    &       &      &     \\
   barrier of 0.25 $eV$ & 0.25  & 4.33 & 1.5 \\
   barrier of 2.0  $eV$ &  2.0  & 4.33 & 1.5 \\
	\hline
 \end{tabular}
 \caption{Potential parameters  used for the inter-molecular interaction. 
The third line shows the parameters of the inner part of the intramolecular reactive bonding potential 
(see appendix \ref{exo_poten_func_form}). 
The same potentials were used for N-C atoms of different molecules.}
 \label{tab:poten_param}
\end{table}

\subsection{Boundary Conditions and Initial Conditions}
\label{sec:init_cond}
We expect the detonation wave to be quasi one-dimensional. 
Hence, the reactive crystal slab  used is chosen to be long and narrow. 
The direction of the propagation axis was chosen as $Z$, 
and it coincides with the [111] crystallographic axis. 
The molecular axes were also oriented along the $Z$ direction. 
Thus, the crystal is very non-isotropic. 
The length of the periodic unit in the [111] direction of an FCC crystal is $\sqrt{6}$ times $r_0$. 
In our case $r_0=7.5 \text{\AA}$, so the length of the periodic unit is $\approx 18.37 \text{\AA}$. 
This distance will be referred to in the following as the unit cell length.
The length of the crystal was  295 unit cells along the $Z$ direction, 
and, in some cases, a crystal with a length of 587 unit cells was used.
Periodic boundary conditions were imposed along the perpendicular directions  $X$ and $Y$. 
To check that the size of the cross section is sufficient we ran simulations of crystals with
various cross sections. 
It was found that detonation properties are converged in the larger cross sections.
The converged cross section which was eventually used consists of 48 molecules in the $XY$ plane. 

The shock wave was initiated by a small crystal pellet, 
assigned with an initial high velocity in the $Z$ direction.
The pellet collided with one of the small faces ($XY$ plane) of the slab.
The velocity of the pellet was not kept constant: 
the MD simulations were of NVE type, 
and the positions and velocities of the pellet's particles were calculated with no distinction.
The pellet was composed of 3 crystal layers along the Z direction 
and had 48 N-C molecules in each layer.
The collision of this pellet was sufficient to initiate a sustainable shock in the primary crystal. 

Most of the simulations were carried with initial temperature of 0$^\circ$K.
A few simulations were carried with higher initial temperatures.

\section{Weak Detonation Waves in the Crystal}

The analysis of weak detonation wave starts from the static properties of the reactive crystal.
The second step describes the initiation process, showing that a stationary detonation wave is formed, independent of initiation process.
The classification as a weak detonation relies on thermodynamics analysis of the phenomena.

The equilibrium properties of the crystal model were characterized. 
The details are summarized in appendix \ref{sec:cryst_charac}. 
A linear relation was found between the velocity of the elastic p-wave
and the stiffness of the potential (the $\beta$ parameter in Eq. (\ref{eq:morse}) ).

\subsection{Initiation of stable detonation waves}

\begin{figure}
 \begin{center}
 \includegraphics*[width=6.0in]{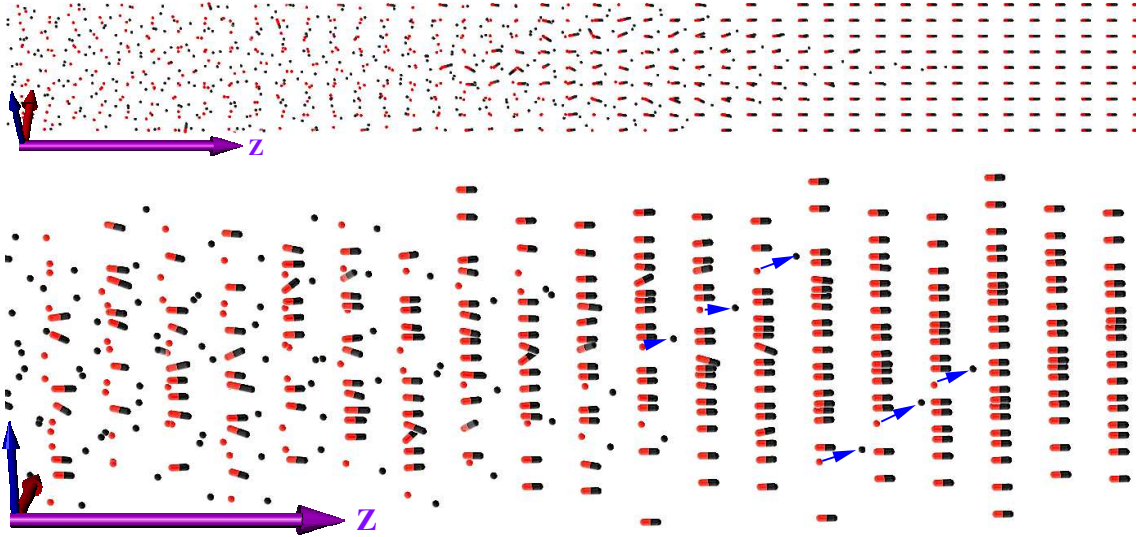}
 \end{center}
 \caption{A snapshot of the detonation wave.
 The propagation axis is the $Z$ direction. Periodic boundary condition are used in the $X$, $Y$ directions.
 The right hand side shows the unperturbed crystal structure.
 The red objects represent the heavy particles (N).
 The black objects represent the light particles (C).
 On the left side of the image one can observe the burnt material after the passage of the shock wave.
 The shock front is characterized by pilot cascades of light particles which are emitted from the decomposed molecules,
 and initiate the next layer in a domino-like effect.
 The lower panel is an enlarged viewpoint of the detonation front.
 The arrows indicate decomposed N-C pairs, corresponding to the pilot cascade.
 \\This simulation was carried out with exothermicity of 6 eV and a barrier of 0.25 eV.
 Eventually, in this simulation, in approximately 20 layers, most of the material is decomposed.
 }
\label{fig:snap}
\end{figure}

We ran NVE simulations of shock waves in the model reactive crystal, initiated by the small pellet described above.
After initiating a shock wave, a transient emerges, eventually stabilizing into  a stable shock front. 
Such simulations were run with different crystal parameters and different initial velocities of the pellet. 
The propagating shock wave initiated a decomposition reaction in crystals composed of reactive molecules. 
When the decomposition reaction kept pace with the shock front, 
the shock wave transformed into a detonation wave. 
The simulations were carried with initial temperature of 0$^\circ$K.
A few simulations were carried with higher initial temperatures, 
in order to verify that the phenomena is not restricted to 0$^\circ$K.

A stable detonation wave is independent of initial conditions.
This was verified by using different initial pellet velocities leading to shock waves that are independent of these initial velocities. 
The details can be found in appendix \ref{sec:impact_indepen}, and in Fig. \ref{fig:diff_init}.

A snapshot of the detonation process is shown in Fig. \ref{fig:snap}.
A distinction between the unperturbed crystal and the burnt material is clearly seen.
An enlarged section of the reaction zone is shown on the bottom.

Similar simulations, with an endothermic intramolecular Morse potential, 
yielded decaying shock waves, with propagation velocity that depends on the 
initial pellet velocity.

\subsection{Classification of the Detonation Waves as Weak Detonations}

After the short transient the shock wave stabilized into a stable weak detonation wave.
The shock front propagated in a very high Mach number, 
practically decoupled from other processes in the crystal. 
The number of molecules that were decomposed by this shock front changed 
from one simulation to another, 
depending on internal potential parameters such as the exothermicity and the 
barrier height for C-N dissociation.
Only a small compression was associated with this front. 
Other properties, such as the mass velocity (Cf. \ref{sec:mas_vel} and the kinetic energy (Cf. \ref{sec:energies}), 
also exhibit an increase in magnitude. 
This shock wave will be referred to as the reaction wave in the following.

The reaction wave travels through an unperturbed crystal,
and therefore its velocity is determined by the crystal properties (cf. Sec. \ref{sec:velo_vs_params}). 
The reaction wave propagation velocity, as well as its amplitude, are independent of the initial impact 
(cf. appendix \ref{sec:impact_indepen}). 
The velocity does depend on the exothermicity of the reaction, 
which is an intrinsic property of the material.
The dependence of the detonation velocity on the reaction exothermicity will be discussed later 
(cf. \ref{sec:exo_satur}).

After the reaction wave, another shock front appeared, 
characterized by a large increase of both the mass velocity and the kinetic energy. 
In the second shock front most of the remaining molecules dissociated, 
leading to a decomposition yield of above 85\%.
This second shock wave will be referred to as the compression wave in the following.
In contrast to the reaction wave, the amplitude of the compression wave decays during its propagation.
The compression wave velocity depends on the impact strength but not on the exothermicity.

The reaction wave is faster than the compression wave: $ u_{reaction} > u_{compression} $
(cf. Figs. \ref{fig:decomp_diff_barr} and \ref{fig:masvelzContour}).
{\em This fact categorizes the reaction wave as a weak detonation: }
The compression wave is a typical shock wave. 
It moves faster than sound waves in the reacted material, 
therefore it sets an upper limit for the sound wave velocity:
$ u_{compression} > u_{sound}$.
Thus: $ u_{reaction} > u_{sound} $.
We see that the reaction wave is supersonic with respect to the matter behind it. 
This is the hydrodynamical definition of weak detonation 
\cite[p. 280]{benartzi2003generalized}. 

\subsection{Thermodynamics Profiles of the Weak Detonation Waves}

The profiles of weak detonations differ from those of normal detonations: 
In weak detonations there is no compression of the material before the reaction.
After the reaction front, 
there is only a small change in the density, pressure, and the velocity of the particles. 

\subsubsection{Different dissociation barrier height}
\label{sec:dif_bar}

\begin{figure*}[htbp]
 \begin{center}
$ \begin{array} {cc}
 \includegraphics*[width=3.0in]{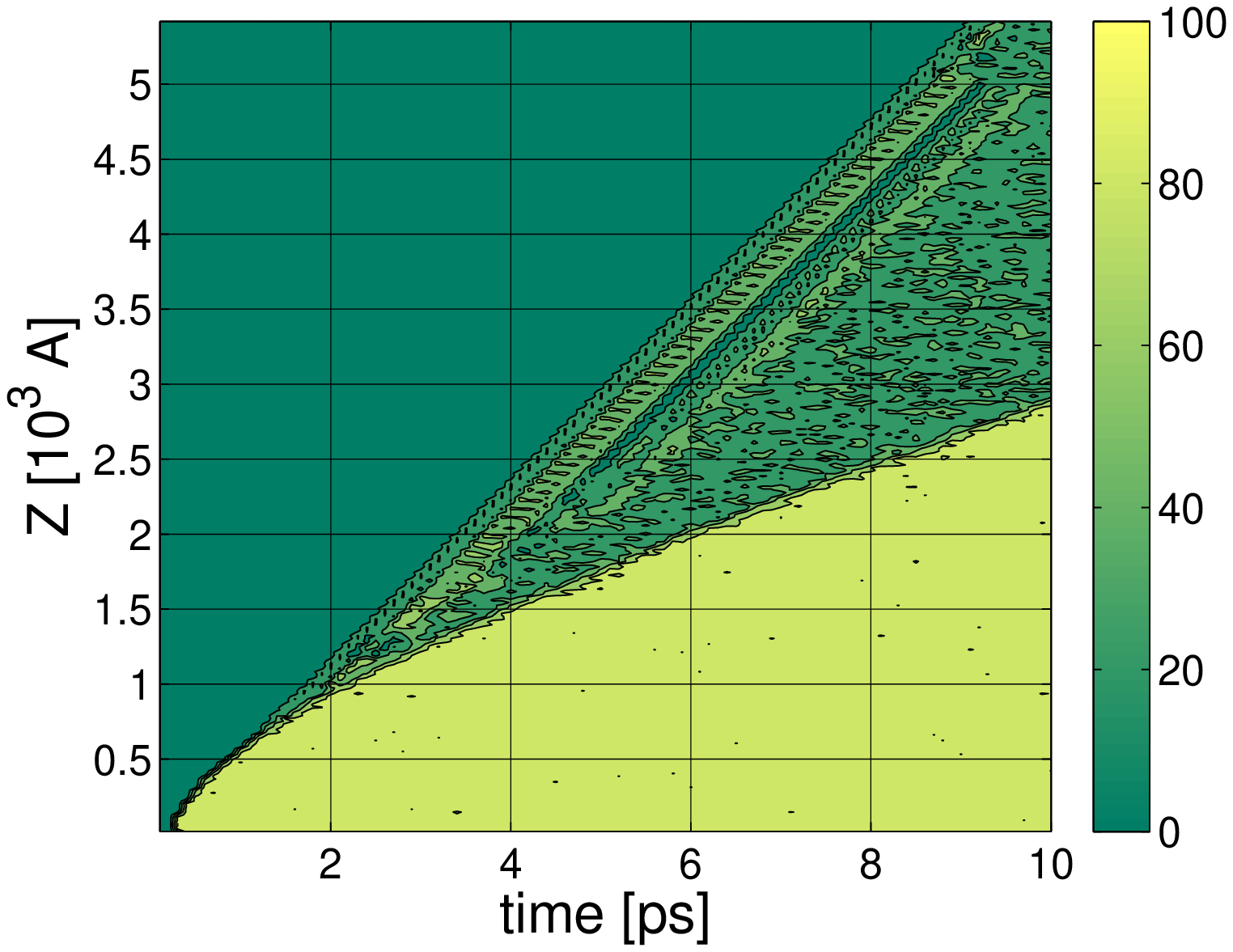} & 
 \includegraphics*[width=3.0in]{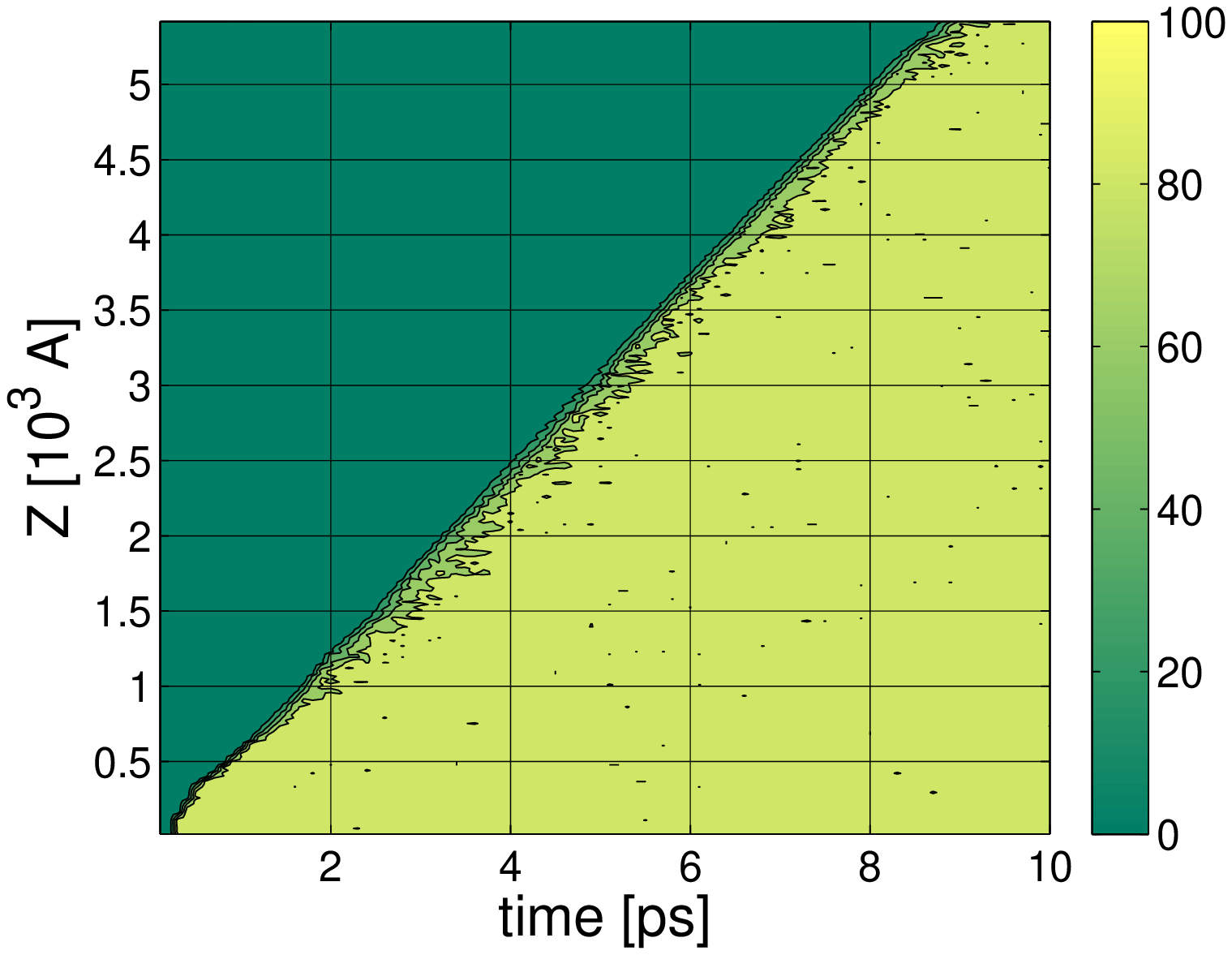} \\ %
  \end{array} $
 \end{center}
 \caption{Contour plots of decomposition percentage during two different simulations.
The left plot is obtained in a simulation with a dissociation barrier of $2\ eV$.
Some of the molecules are decomposed by the reaction wave. 
Most of the other molecules (around 90\%) are decomposed by the compression wave.
The right plot is obtained in a simulation with a decreased dissociation barrier of $0.25\ eV$.
More than 85\% of the molecules are decomposed by the reaction wave.
In both cases the exothermicity is $6\ eV$,
and the initiating pellet velocity is $\sim90\ km/s$.
\\ The reaction front velocity in the case of the higher barrier, $\sim60\ km/s$, 
is very similar to the case of the smaller one, , $\sim62.5\ km/s$.
}
 \label{fig:decomp_diff_barr}
\end{figure*}

Fig. \ref{fig:decomp_diff_barr} compares the decomposition fraction of two simulations 
as a function of time and position.
The first corresponds to molecules that have an N-C dissociation barrier of $2\ eV$ (left plot). 
The second corresponds to molecules with a dissociation barrier height of $0.25\ eV$.
The reaction exothermicity in both simulations is $6\ eV$.
For a dissociation barrier of $2\ eV$, 
$\sim$ 35\% of the molecules are decomposed by the reaction wave.
When the compression wave passes through the partially ``burnt'' material, 
most of the remaining molecules decompose
(small fluctuations around 90\%).
For a smaller dissociation barrier, $0.25\ eV$, 
more than 85\% of the molecules are decomposed simultaneously with the \textit{reaction} shock front. 

The reaction wave velocities are almost independent of the dissociation barrier height,
as can be seen in Fig. \ref{fig:decomp_diff_barr}.
The barrier does influence the mass velocity and the decomposition fraction.
The decoupling of the reaction wave from the compression wave and from the dissociation process 
results from the fact that this is the fastest wave in the material, leaving behind the slower processes.
This phenomena is another signature of a weak detonation wave.

\subsubsection{Mass velocity profile}
\label{sec:mas_vel}

\begin{figure}[htbp]
 \begin{center}
   \includegraphics*[width=3.0in]{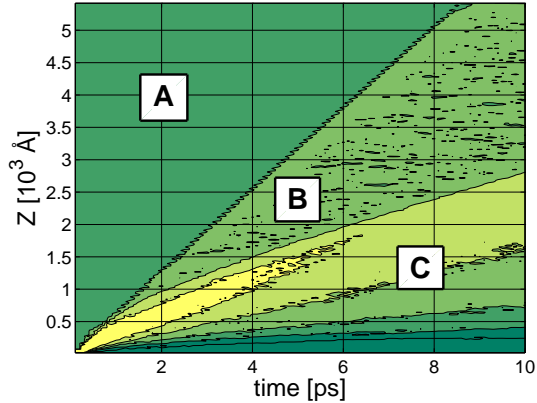}
 \end{center}
 \caption{Contour plot of mass velocity along the $Z$ direction.
The dissociation barrier height in this simulation was $0.25\ eV$, 
and the reaction exothermicity was $6\ eV$.
The initiating pellet velocity was $\sim90\ km/s$.
Three regions are marked in the plot: 
Region A is the pre-shocked material. The mass velocity here is $0$.
Region B is the material after the reaction wave. 
The mass velocity here fluctuates around an average value of $0.3\ km/s$ (see Fig \ref{fig:masvelzANDx} for more detailed profiles) .
The mass velocity plot shown here can be compared to the decomposition percentage 
during the same simulation, which is shown on the right plot of figure \ref{fig:decomp_diff_barr}:
The decomposition front and the discontinuity of the mass velocity coincide, 
characterizing the reaction wave.
Region C is the material after the compression wave. 
The mass velocity on the peak of the compression wave decays from around $20\ km/s$ at 1 ps to $8.5\ km/s$ at 10 ps.
The compression wave is visible on the mass velocity plot,
but is almost absent on the decomposition plot: 
For this barrier height, most of the molecules have already decomposed before the compression wave reached them.
}
 \label{fig:masvelzContour}
\end{figure}

The mass velocity characterizes the mass current, is defined as 
$ \left\langle v \right\rangle = \frac{\sum_i m_i v_i}{\sum_i m_i} $.
Figure \ref{fig:masvelzContour} shows a contour plot of the $Z$ component 
of the mass velocity during the simulation. 
The mass velocity plot shown can be compared to the decomposition percentage 
during the same simulation (right hand side of Fig.  \ref{fig:decomp_diff_barr}):
The decomposition front and the discontinuity of the mass velocity coincide, 
and define the reaction wave.
The compression wave is observed on the mass velocity plot,
but is absent from decomposition plot 
since the majority of the molecules have already decomposed.

\begin{figure}[htbp]
 \begin{center}
$ \begin{array} {c}
   \includegraphics*[width=3.0in]{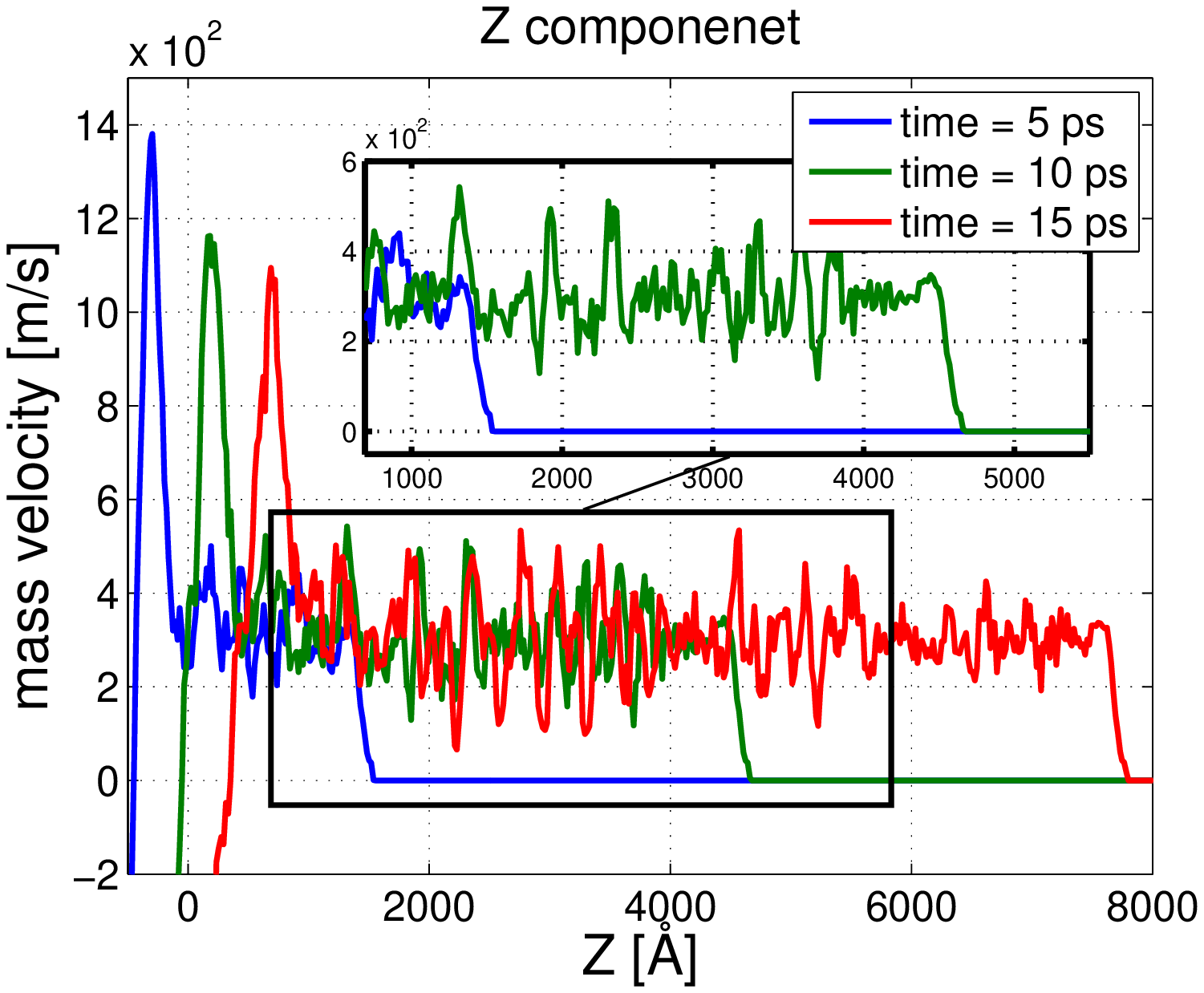} \\
   \includegraphics*[width=3.0in]{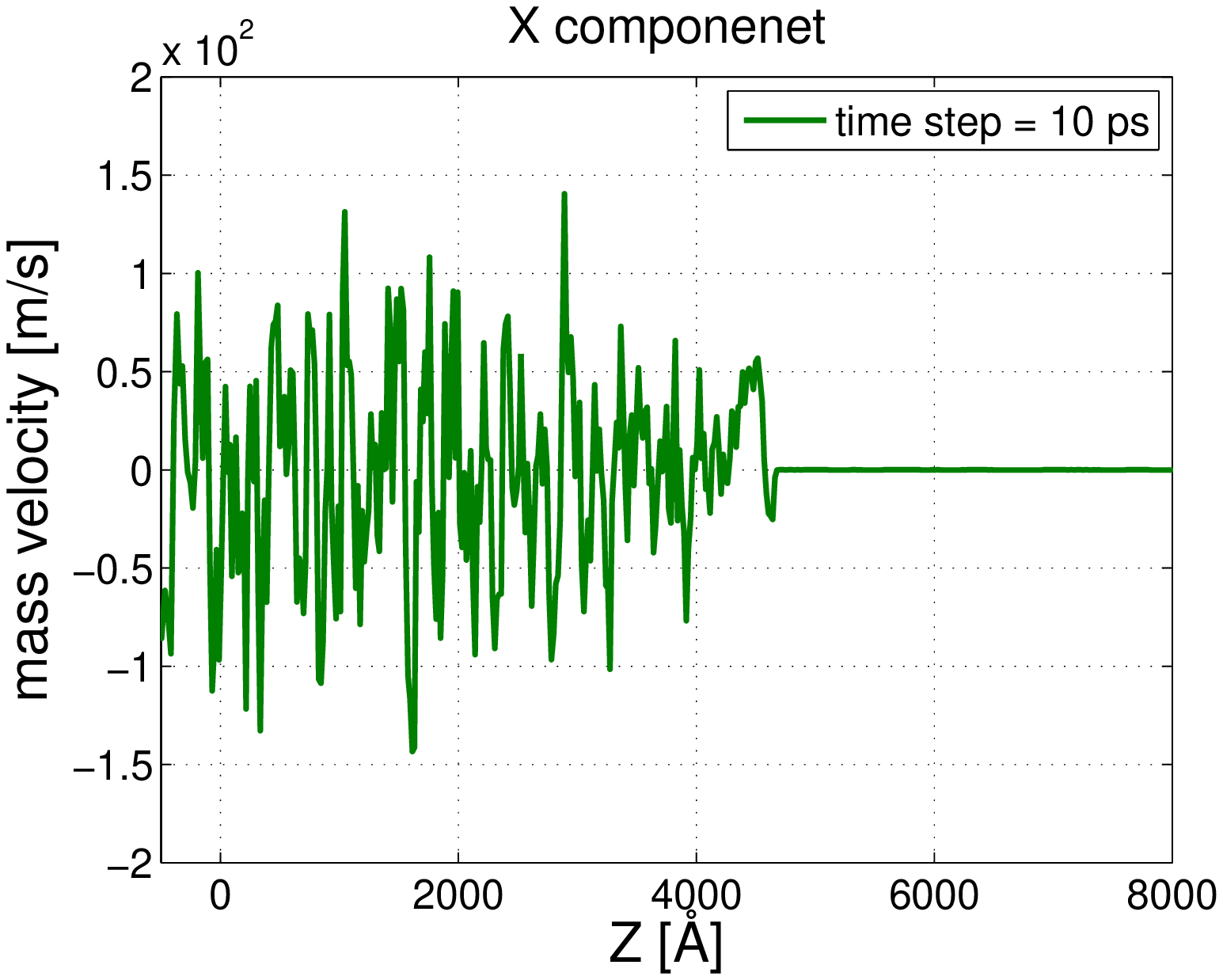} \\
  \end{array} $
 \end{center}
 \caption{Mass velocity profiles along $Z$  and $X$ directions. 
The exothermicity of the reaction in this simulation was $6\ eV$, and the barrier was $0.25\ eV$.
The values were determined by averaging over bins with constant 96 particles in each bin.
The left edge of the crystal is placed at -500 \AA. 
The top plot shows the profile of the mass velocity in the $Z$ direction, at three time values.
The bottom plot presents the mass velocity along the $X$ direction at a single time value.
Along the Z direction (top figure), 
the largest  amplitude (the left peak on each time step) is associated with the secondary compression wave. 
At later times the peak mass velocity decays due to dissipation. 
The first shock front (the right peak on each time step) is associated with the reaction wave.
The reaction wave amplitude does not decay.
The insert shows a zoom of the first shock fronts for the two shorter times.
The average mass velocity after the reaction front, $\sim0.3\ km/s$, is much smaller than the reaction front velocity, $\sim65\ km/s$.
}
 \label{fig:masvelzANDx}
\end{figure}

Figure \ref{fig:masvelzANDx} shows profiles of mass velocity at different snapshots during the simulation. 
The values were averaged over bins with a constant number of particles, 96, in each bin.
The top plot shows the $Z$ component of the mass velocity along the solid for three time values.
The largest amplitude (the left peak for each time) is associated with the compression wave. 
At later times the peak mass velocity decays due to dissipation. 
The first shock front (the right sudden change in velocity for each time) is associated with the reaction wave.
The insert shows a zoom-in on the reaction fronts.
The average mass velocity after the reaction front, $\sim0.3\ km/s$, is much smaller than the reaction front velocity, $\sim65\ km/s$.
This is consistent with a small increase of density after the reaction front (Cf. Fig. \ref{fig:potentDensProfile}).
It is clear that the reaction wave amplitude does not decay.
When comparing simulations with different parameters, we find that the reaction wave amplitude 
depends on the reaction exothermicity and is independent  of the impact strength.
As discussed above, in the introduction and in section \ref{sec:impact_indepen}, 
this is a characteristic of a detonation wave.
The compression wave behaves differently: 
it has a decaying amplitude that does not depend on the exothermicity 
but does depend on the impact strength.
The compression wave is a regular compression  shock wave.
It travels through an unstructured matter 
composed of reaction decomposition products.

The $X$ component of the mass velocity is shown in the bottom plot for a single snapshot.
The shock front is characterized with a abrupt onset of fluctuations.
There is a minor change after the first shock front. 

\begin{figure}[htbp] 
 \centering \includegraphics*[width=3.0in]{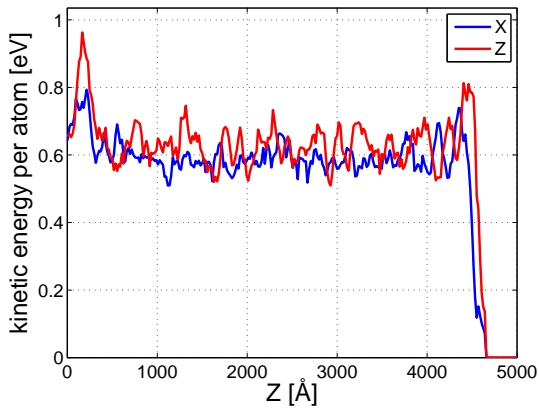}
  \caption{Normalized kinetic energy profile at time = 10 ps.
The exothermicity of the reaction in this simulation was $6\ eV$, 
and the dissociation barrier was $0.25\ eV$.
The values were averaged over bins with constant number of particles, 96, in each bin.
The blue line corresponds to kinetic energy along the X direction, and the red line to kinetic energy along the Z direction.
The large peak on the right corresponds to the reaction wave front.
}
 \label{fig:kineticProfile}
\end{figure}

\subsubsection{Energies and temperature}
\label{sec:energies}

\begin{figure}[htbp]  
 \centering \includegraphics*[width=3.0in]{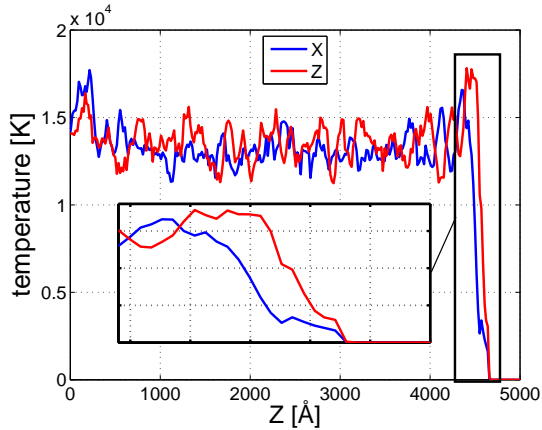}
  \caption{Local temperature profile at time = 10 ps.
The parameters are the same as in Fig \ref{fig:kineticProfile}.
The blue line represents temperature along the X direction, and the red line along the Z direction.
The insert shows an unthermalized region:
The large peak on the right is the reaction wave front.
The increase of the $Z$ component of the local temperature precedes the increase of the $X$ direction.
}
 \label{fig:temperatureProfile}
\end{figure}

\begin{figure}[htbp]  
 \centering \includegraphics*[width=3.0in]{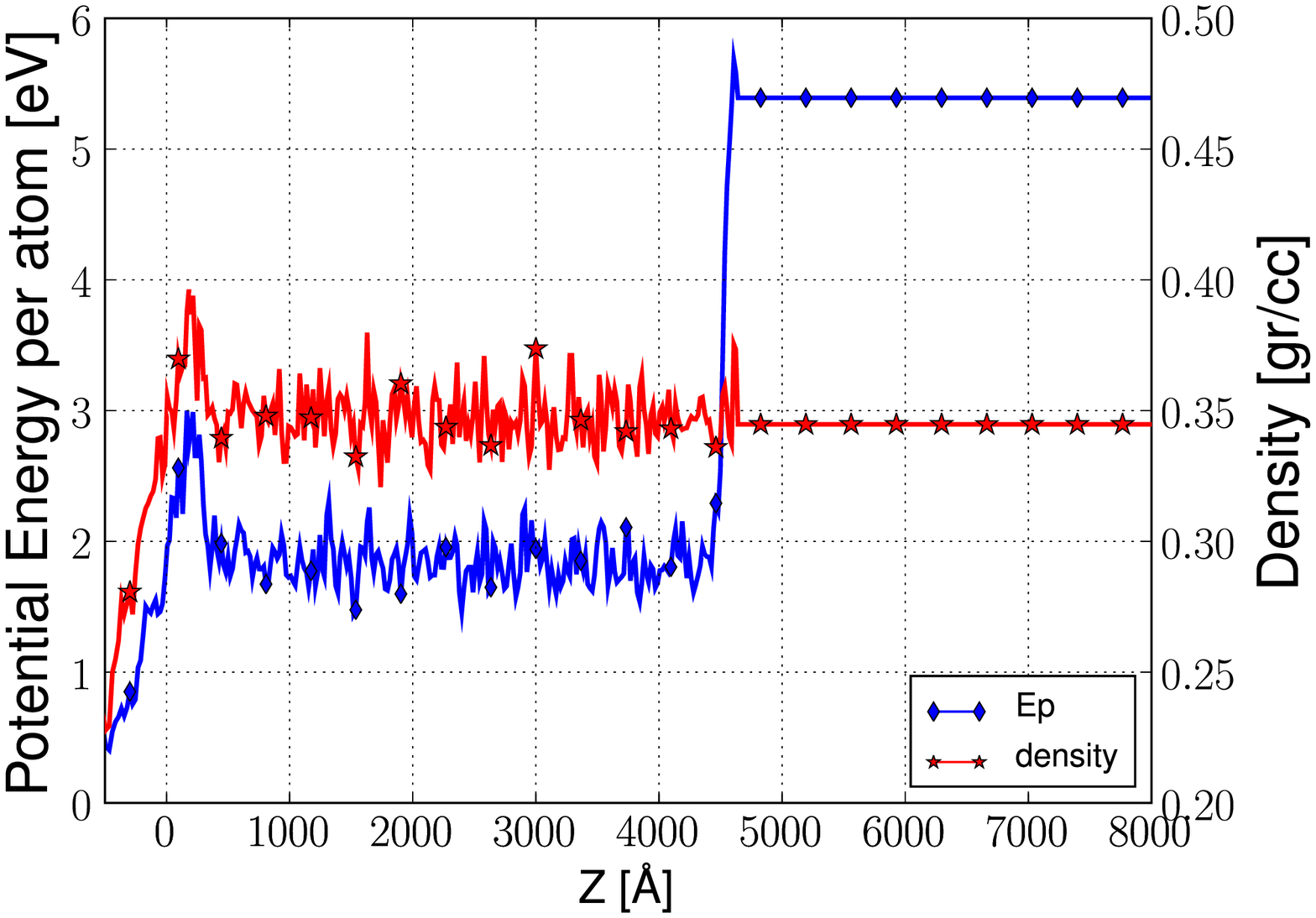}
  \caption{Potential energy (Ep) and density profiles at t = 10 ps.
The parameters are the same as in Fig \ref{fig:kineticProfile}.
The first shock front displays a small increase of the potential energy, 
due to a small compression before the dissociation.
This small increase is followed by a large decrease of the potential energy, 
due to the energy release of the exothermic decomposition.
The average density after the reaction front is not significant higher than the density before this front.
The peak on the left is associated with the compression wave.
}
 \label{fig:potentDensProfile}
\end{figure}

Further insight on the detonation process can be obtained by inspection of the 
kinetic energy and the local temperature profiles, 
shown in figures \ref{fig:kineticProfile} and \ref{fig:temperatureProfile}, respectively.
The profiles correspond to an intermediate time of 10 ps.
The local temperature is defined as 
$ T_l = \frac{1}{N k_B} \sum_i^N m_i \left( v_i - \left\langle v \right\rangle \right) ^2 $,
where $ \left\langle v \right\rangle $ is the mass velocity defined above.
The values were determined by averaging over bins with a constant (96) number of particles.
The reaction wave front can clearly be seen as a sharp change in kinetic energy and temperature. 
The insert in the local temperature figure 
shows a small region after the first shock front which is not thermalized. 
The increase of the $Z$ component of the local temperature precedes the corresponding 
increase of the $X$ component.

Figure \ref{fig:potentDensProfile} 
shows the potential energy and the density profiles at the same time step.
On the right, a small increase of the potential energy is observed, 
followed by a large decrease.
The small increase is caused by a minor compression before the dissociation starts, 
as seen in the density profile.
The large decrease that follows is caused by the energy release of the 
exothermic decomposition process.
The average density after the reaction front is only slightly larger than the density before this front.
This is consistent with the small increase in the mass velocity (Cf. Fig. \ref{fig:masvelzANDx}).
This phenomena characterizes weak detonations.

\section{Saturation of the weak detonation velocity}
\label{sec:velo_vs_params}
The detonation wave can be characterized by a stable detonation velocity independent of initiation parameters,
as discussed in appendix \ref{sec:impact_indepen}.
The terminal velocity of the wave is a function of the microscopic parameters used in the model.
Insight aimed at deciphering the phenomena is obtained by a systematic study of the
variation of the detonation velocity as a function of the parameters of interest.

\subsection{Intramolecular potential parameters: exothermicity of the reaction}
\label{sec:exo_satur}

\begin{figure}[htbp]  
 \centering \includegraphics*[width=3.0in]{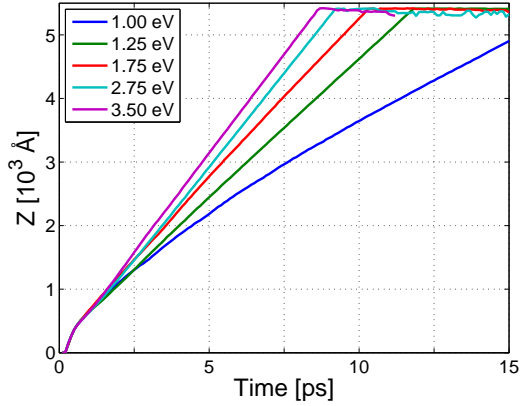}
 \caption{Reaction front propagation in crystals with different exothermicity values. 
All other parameters are identical.
}
 \label{fig:diff_E}
\end{figure}

\begin{figure}[htbp]  
 \centering \includegraphics*[width=3.0in]{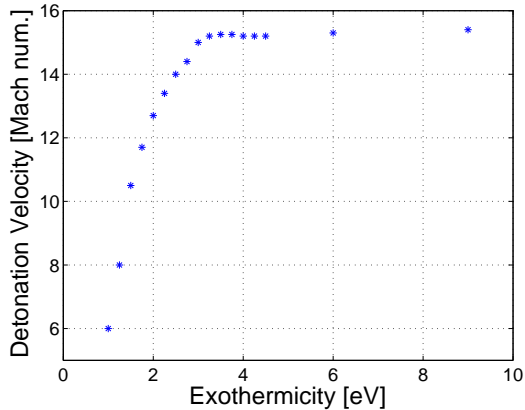}
 \caption{
Detonation velocity in the crystal as a function of reaction exothermicity. 
The detonation velocity is measured in reduced units, normalized to the p-wave velocity, which is $4.6\ km/s$.
We can see the increase of detonation velocity as a function of exothermicity 
at low exothermicity values, until approximately 3 eV.
The saturation of the detonation velocity at high exothermicity values 
can also be seen.
The saturation was verified by simulations with very high exothermicity values: 
there is a very small increase of the detonation velocity in crystals with
higher exothermicity values (12 eV, 15 eV, 18 eV, and even 27 eV. 
These results are not shown here).
}
 \label{fig:velo_exo}
\end{figure}

Simulations of detonation using reactive slabs were carried out with
different values of the exothermicity. 
All other parameters, such as crystal structure, impact magnitude, intermolecular potential, 
particles masses, initial conditions, were kept constant. 
Figure \ref{fig:diff_E} displays the reaction propagation, which is determined by the dissociation front, 
in crystals with different exothermicity values. 
In the exothermicity range displayed in the figure ($1.0-3.5\ eV$), 
the detonation velocity increases as a function of exothermicity. 
At larger exothermicity values, the detonation velocity reaches saturation:
simulations with exothermicity values above $3.0\ eV$, 
show only a minor increase in the detonation velocity.
Figure \ref{fig:velo_exo} displays the variation in detonation velocity as a 
function of reaction exothermicity. 
The kinetic energy of the particles behind the reaction wave was found to depend linearly 
on the exothermicity of the reaction. Thus it is responsible for the energy balance.
Nevertheless, in weak detonations this additional kinetic energy is decoupled from the reaction shock front.

The saturation effect is an interplay between three factors:
the crystal rigidity, responsible for the non-linear wave propagation,
shear effects, responsible for energy dissipation at the reactive shock front,
and kinematic effects, which determine the partitioning of the energy release
during the decomposition of a single molecule.

Some insight regarding kinematic effects and the crystal rigidity is discussed in the following, 
cf. sections \ref{sec:kinematics} and \ref{sec:rigidity}.
The non-linear effect of shear was not explored systematically in this study.
The convergence of the calculations was checked to an increase in the XY cross section (cf. \ref{sec:init_cond}).
While checking convergence, we found that increasing the cross section decreases the detonation velocity.
This is an indirect indication that the detonation velocity is sensitive to shear.

\subsection{Kinematic effects}
\label{sec:kinematics}
\begin{figure}
 \centering \includegraphics*[width=3.0in]{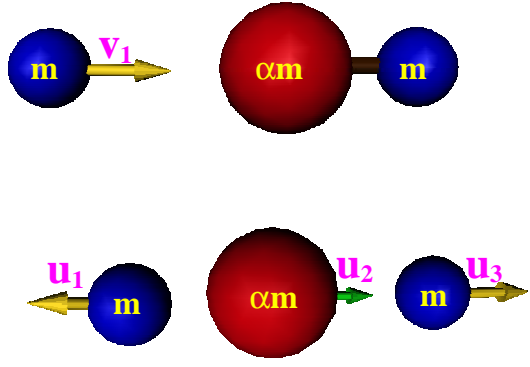}
 \caption{A molecule is hit by a light particle, emitted from neighboring decomposition (top diagram).
After the collision, the molecule decomposes (bottom diagram).}
 \label{fig:1Dcrash}
\end{figure}

In weak detonations the decomposition mechanism of the reactive molecules has to be linked 
to the characteristics of the detonation wave.
A propagation by a domino-like effect, as seen in Fig. \ref{fig:snap}, 
requires the decomposition to be asymmetric with respect to the $Z$ direction. 
Insight is obtained by examining a simplified 1-D model of decomposition of a single diatomic molecule.
Initially the molecule is at rest. 
A light particle emerges from the decomposition of the neighboring molecule,
and hits the molecule from the heavier side (Fig. \ref{fig:1Dcrash}, top).
The collision initiates the decomposition process.
At the end of the decomposition, the light particle of this molecule is emitted toward the next molecule 
(Fig. \ref{fig:1Dcrash}, bottom).
The mass of the light particle is denoted by $m$, and the mass of the heavier particle is $\alpha m$ ($\alpha>1$). 
After the initiation collision, the molecule decomposes and the heat of the reaction, $E$, releases as kinetic energy. 
The velocity of the colliding particle is denoted by $v_1$. 
After the decomposition, the velocities are denoted by $u_1, u_2, u_3$, for the
colliding particle, the heavy particle, and the light particle, respectively.

Under these conditions the equations for momentum and energy conservation become:
\begin{eqnarray}
\begin{array}{l}
m v_1=m u_1+ \alpha m u_{2} + m u_3 \\
\frac{1}{2} m v_1^2+E = \frac{1}{2} m v_1^2 + \frac{1}{2} \alpha m u_2^2 + \frac{1}{2} m u_3^2
\end{array}
\label{eq:direct_decomp}
\end{eqnarray}
where we assume that the collision is complete, meaning that all potential energy has been consumed. 
In steady state detonation, the initial velocity of the colliding  particle should be equal to the velocity 
of the emitted  light particle:
\begin{equation}
 v_1=u_3,
\label{eq:steady_state}
\end{equation}
and this should also become the group detonation velocity.

Two mechanisms come to mind: a direct and a delayed decomposition.
In the direct decomposition mechanism, Eq. (\ref{eq:direct_decomp}) and (\ref{eq:steady_state})
give the complete kinematic picture. 
Substituting Eq. (\ref{eq:steady_state}) into Eq. (\ref{eq:direct_decomp}) 
cancels out the terms $v_1$ and $u_3$. 
Therefore, the detonation velocity is undetermined without further assumptions.
In order to determine the wave propagation velocity,
more details on the dissipation mechanism, the crystal structure and the non-linear properties are needed.

Next, we analyse the delayed decomposition mechanism.
This mechanism has two steps: 
first, the colliding particle has an elastic encounter with the molecule. 
The molecule after the collision will acquire a velocity of $u_{23}$ and mass of $(\alpha + 1)m$. 
On the second step, the molecule will dissociate into its components. 
The equations for momentum and energy conservation of the first step are:
\begin{eqnarray}
\begin{array}{l}
mv_1 = mu_1 + (\alpha + 1)mu_{23} \\
\frac{1}{2} mv_1^2 =\frac{1}{2} m u_1^2 +\frac{1}{2} (\alpha +1) m u_{23}^2
\end{array}
\end{eqnarray}
And the equations for momentum and energy conservation of the second step are:
\begin{eqnarray}
\begin{array}{l}
(\alpha +1) mu_{23} = \alpha m u_2 + mu_3 \\
\frac{1}{2} (\alpha +1) m u_{23}^2 + E = \frac{1}{2} \alpha m u_{2}^2 + \frac{1}{2} mu_3^2
\end{array}
\end{eqnarray}
Again, we require the steady state condition, Eq. (\ref{eq:steady_state}).
Solving these equations 
shows that the velocities $v_1=u_3$, which are equal to the detonation velocity, 
are proportional to $\sqrt {E/m}$. 

A strong detonation process is adequately described by a delayed mechanism,
since  there is a delay between the compression (which leads to collisions) and the decomposition. 
The $\sqrt {E/m}$ dependence of the strong detonation velocity is in accordance with the hydrodynamical theory.
This $\sqrt{E}$ dependence
was obtained in Ref. \cite{heim2007potential_influence} in MD simulations of detonations in the REBO model,
and in Ref. \cite{fomin2006Molecular_Characteristics_Influence} in another crystal model.
A different result was obtained in an earlier study \cite{elert1998variation_Energy_Release}, 
in which a linear dependence of the detonation velocity on the energy released is reported.

A weak detonation process is adequately described by the direct mechanism
since the reaction coincides with the shock front. 
In weak detonations, we find the detonation velocity to saturate, becoming independent of the energy release. 
This means that in our 1-D model, only the results of a direct decomposition mechanism can fit this phenomena.
The relation of direct reaction mechanism to weak detonations was shown also in 
hydrodynamic analysis of the ignition stage \cite{Dold1991Comparison_shock_initiations}. 

The mechanism discussed above suggests a difference 
between the two possible reaction wave propagation directions: 
Different behavior is expected if the reaction wave emits the heavier particle of the molecule, 
which in turn collides with the lighter particle of the neighboring molecule.
This anisotropic behavior of the reaction wave was examined by initiating the shock wave in the opposite edge,
impacting from the lighter side of the molecules. 
Under these conditions, no stable constant velocity reaction waves were formed.

\begin{figure}[htbp]
 \centering \includegraphics*[width=3.0in]{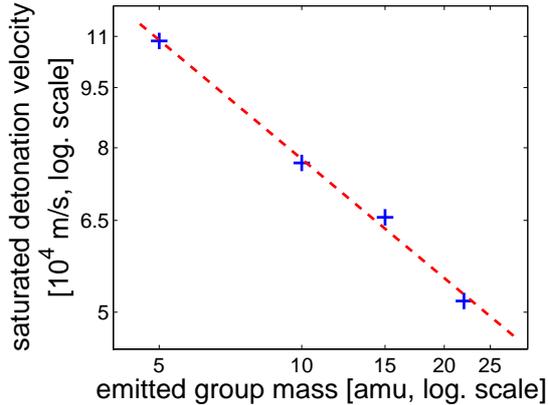}
 \caption{Saturation values of detonation velocities, marked with asterisks, 
as a function of the mass of the emitted $C$ group.
The graph is plotted in logarithmic scale on both axis.
In all cases, this saturation value is obtained when the exothermicity is close to $3\ eV$.
A linear fit of the data in the log-log scale is also plotted.
The slope of the linear fit is close to $-0.5$ ($-0.49$), 
indicating that the detonation velocity is proportional to $1/\sqrt{m}$.
}
\label{fig:diff_mass}
\end{figure}

The dependence of the detonation wave velocity on the mass of the emitted particle was also examined.
We performed simulations of detonations where the mass of the lighter particle was varied 
while maintaining the same structure and force field. 
In order to keep the total mass unchanged, the mass of the heavier particle was also changed. 
The masses used were 22, 10 and 5 amu for the $C$ particle (and 40 ,52 and 57 amu for $N$, respectively). 
Several simulations were performed, with different exothermicity values.
We expect that decreasing the mass of the $C$ lighter particle  will result in a higher velocity, hence faster detonation. 
This has been already suggested by Peyrard et al. \cite{peyrard1985mmc} in a 1-D model. 
The magnitude of exothermicity at which the detonation velocity saturated was found in all cases to be close to 3 eV. 
However, the saturation velocity depended on the mass of the emitted $C$ particle. 
Lower mass of this particle led to an increased detonation velocity. 
These results are presented in Fig. \ref{fig:diff_mass}, in log-log scale. 
A linear fit to the velocity logarithm as a function of the mass logarithm was calculated 
and is displayed in the graph.
The resulting slope is close to $0.5$ ($0.49$), 
which reveals a $1/\sqrt{m}$ dependence of the detonation velocity on the mass of the emitted group.
This dependence is identical to the dependence of the shock velocity of solitons in a Toda lattice.

\subsection{Crystal rigidity: Intermolecular potential parameters}
\label{sec:rigidity}
Simulations of the propagation of small amplitude displacements in the crystal showed that the elastic 
P-wave velocity depends linearly on the $\beta$ parameter of the intermolecular Morse potential
(Eq. (\ref{eq:morse})), i.e. the stiffness.
This is in agreement with the propagation of elastic waves in a 1-D crystal model.
See appendix \ref{sec:pwave_vel} and Fig. \ref{fig:p_bet} for details.

\begin{figure}[htbp]
 \centering \includegraphics*[width=3.0in]{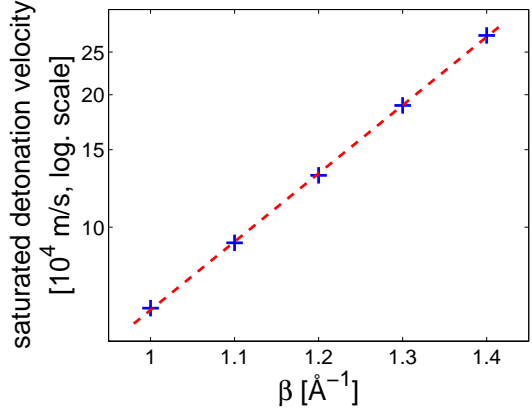}
 \caption{Comparison of reaction propagation velocities (scaled logarithmically) in 5 
different crystals, $\beta$ ranges from $ 1\ \text{\AA}^{-1} $ to $ 1.4\ \text{\AA}^{-1} $ 
(when $\beta$ is the coefficient inside the Morse potential's exponent).
The exponential dependence suggests that the saturation of the detonation velocity is 
caused by the repulsive part of the intermolecular potential. 
This is different from the elastic p-wave velocity dependence, 
which is governed by the behavior of the potential near the equilibrium point 
(see appendix \ref{sec:pwave_vel} and Fig. \ref{fig:p_bet}). 
} 
 \label{fig:det_bet}
\end{figure}

The shock wave velocity is crucially dependent on the rigidity of the crystal, 
which is determined by $\beta$. 
To explore this dependence we carried out simulations of detonations in crystals with different $\beta$ values. 
The first effect observed is that an increase in $\beta$ leads to an increase in 
magnitude of the exothermicity for which the saturated reaction wave velocity is observed. 
The second effect found is that the saturation reaction wave velocity increased exponentially as a function of $\beta$.
Figure \ref{fig:det_bet} shows the logarithm of the saturated reaction wave 
velocity as a function of $\beta$. 
The exponential dependence can be clearly seen.
The exponential dependence suggests that the saturation of the reaction wave velocity 
depends on the repulsive part of the intermolecular potential. 
This is different from the elastic p-wave velocity which  is governed by the shape of the potential near equilibrium
and varies linearly as a function of $\beta$.

To verify that the saturation of the reaction wave velocity 
is governed by the repulsive part of the intermolecular potential, 
we ran additional simulations.
In these simulations we varied the stiffness  of the inner part of the repulsive potential 
without altering other parameters of the crystal: 
The potential was combined from a repulsive part of Morse potential with $\beta = 1.2\ \text{\AA}^{-1} $
and an attractive part of Morse potential with $\beta = 1.0\ \text{\AA}^{-1} $. 
The reaction shock wave velocities in these crystals were compared to the previous simulations. 
We found that the reaction waves velocities  were almost identical to those with  
$\beta = 1.2\ \text{\AA}^{-1} $.
To conclude the reactive shock-wave velocity is governed by the repulsive part, and independent 
of the soft long range part of the potential. This phenomena
characterizes solitary waves, as has been suggested by Toda \cite{Toda1976,Toda1983} and 
by Rolfe et al. \cite{rolfe1979solitons}.

\section{Discussion}

The purpose of this study was to bridge the gap between a first principle microscopic model 
of detonation and bulk hydrodynamical theories. 
The first task was to obtain a stable detonation wave independent of initial conditions.
For this task we constructed a reactive molecular crystal model characterized by pair potentials.
The equilibrium properties of the crystal are typical. 
It is stable at low temperatures and melts at temperatures which scale with the binding energy. 
The model crystal also possesses linear elastic waves.
The detonation potential was added by making the molecule metastable to 
dissociation releasing a significant amount of energy. 
The model fulfilled the expectations and a stable detonation wave was identified. 

Further analysis,  which compared the results obtained in the simulations to hydrodynamical theory, 
revealed a puzzling picture. 
The detonation wave did not have the characteristics of the common solutions of the ZND model. 
Some of the differences are:
\begin{itemize}
\item{The compression at the shock front was minor.}
\item{The chemical reaction coincided with the shock front.}
\item{The shock velocity was supersonic with respect to the burnt material left behind.}
\end{itemize}
Searching for a meeting point with hydrodynamical theory, 
we conclude that the phenomena we observed in the MD simulation is a {\em weak detonation}.   
Weak detonation is a possible hydrodynamical solution
in which the shock wave is supersonic with respect to the material left behind. 
The characteristics of weak detonations are different from normal detonations:
The weak detonation velocity is higher,
and the pressure after completion of the reaction is smaller than in the normal detonation case.
Also, in weak detonations there is no compression of the material before the reaction.
Zel\'dovich argues that such solutions are usually unattainable for substances that are initially inert 
\cite{zeld1942distribution}.
However, he points out that weak detonations might occur if the chemical reaction would start in the 
initial state without preliminary heating of the substance by the shock wave \cite{zeldovich1944Theory}. 
Dremin states that if self-ignition occurs at a pressure lower than the CJ pressure, a weak detonation wave is observed 
\cite{dremin2000discoveries}.

Why are weak detonations attainable and stable in our case?  
The answer is in the kinematic behavior of the crystal. 
The shock propagates with characteristic of a nonlinear solitary wave. 
The propagation velocity is determined by the repulsive part 
of the interatomic potential which is similar to solitons on a Toda lattice. 
The other kinematic property is the mass of the group that emitted 
from the dissociating molecule.
When the shock front comes from the heavy side, the light particle is breaking out
with the majority of the kinetic energy, initiating the reaction on the next crystal plane.
This can be imagined as a shock propagation by the domino effect which never looks back. 
The ideal domino effect has no change in the density after the front. 
This is accompanied by a negligible increase in the mass velocity.
The melting of the crystal and the equilibration of the burnt material lag behind the shock 
front and are decoupled from it due to the supersonic velocity.
Remarkably, when the molecule orientation is the opposite (heavy particle is placed ahead 
of the light one in the shock direction) then the detonation does not reach a steady state.
It explains the details of the solitary wave ignition mechanism. If the light particle
is pushed forward at the high (supersonic) velocity after the decomposition, 
it quickly hits another molecule in the next crystal plane, allowing the reaction 
to propagate at supersonic speed due to the domino effect.
On the contrary, if the heavy particle is placed ahead, it breaks out at much slower velocity,
delaying the propagation of the decomposition reaction.

These observations pose additional questions: 
\begin{itemize}
\item{What are the conditions that are necessary to observe weak detonation in experiments?}
\item{What are the conditions that an MD simulation will reconstruct the standard solutions of the ZND model?}
\end{itemize}
It seems that a  prerequisite for a stable weak detonation is a stiff molecular crystal which supports 
fast propagation of nonlinear shock waves. 
In addition, the crystal should be very non-isotropic. 
The molecules are oriented with the light particle pointing toward the propagation direction. 
We have carried out preliminary simulations with triatomic molecules with similar effects. 
Anisotropic shock initiation sensitivity has been observed in detonation of explosives composed of 
single molecular crystals \cite{yoo2000anisotropic}.
Recent calculations reveal this anisotropic sensitivity 
\cite{conroy2008anisotropPETN, Conroy2009nitromethane}.
Weak detonations were shown to occur in the shock initiation process in the case of inhomogeneous explosives 
or inhomogeneous initial conditions.
\cite{Zeldovich1980nonuniform_initial_conditions, Bdzil1992weak_to_ZND, kapila2002temperature_gradient}. 
Weak detonations in mixtures which have nonmonotonic energy release were shown experimentally 
\cite{balalaev2008weak_detonation_experiment}. 
It has been demonstrated that a quasi-steady form of weak detonation plays an integral role in describing
shock-induced transition to detonation in an explosive material 
\cite{short2002weakstrong}.
The transition to normal detonation was shown to occur effectively at the point where the weak
detonation slows to the CJ velocity. 
In cases of very porous materials it can happen that the decomposition wave 
will remain faster than the compression wave, 
stabilizing the weak detonation solution \cite{Tarver1982chemicalEnergyRelease}.
Finally, a weak detonation requires a decomposition reaction which can follow pace with the shock front even
at relatively low temperatures.
This dictates a time scale of a few tens of femtoseconds.  

The prerequisite for the MD simulation to generate the 
standard results of the Chapman-Jouguet or the more elaborate ZND model is
a more complex and slower chemical reaction which can justify the quasi-equilibrium assumption.
This influence of the speed and complexity of the reaction can be seen even in 1-D models:
Elert et al. \cite{elert19891D} used three body interaction potentials in a 1-D model,
and got a stable detonation without introducing artificial frictional forces.
The studies in Ref.  \cite{zybin2010PETNsens} using realistic force fields aim at this direction.
Nevertheless, the quasi-equilibrium assumption has been criticized \cite{landerville2009PETN}.
In the present study we find that most regions of the detonating crystal are well described by 
quasi-equilibrium assumptions, except the vicinity of the shock front.

\section*{Acknowledgments}
We would like to thank Matania Ben-Artzi, for insightful and fruitful discussions.
We also want to thank Naomi Rom and Ido Schaefer for helpful discussions.
The study was partially supported by the Center of Excellence for Explosives Detection, 
Mitigation and Respons, DHS.

\appendix

\section{MOLDY - The Molecular Dynamics Program}
\label{sec:moldy}

The basic simulation program used  to integrate the molecular dynamics equations was MOLDY \cite{Refson2000310}. 
MOLDY is suitable for the present purpose due to two primary reasons:
\begin{itemize}
 \item{
The equations of motion are integrated in MOLDY using a modified version of the Beeman algorithm \cite{Beeman1976130, Refson2000310}. 
During detonation simulations
there is rapid energy exchange between potential and kinetic energy.
Simulations carried out with the Verlet algorithm failed to conserve
energy properly. The Beeman algorithm, with higher accuracy in
velocities, was found to be adequate.   }
 \item{
The neighbor list in MOLDY is built using the linked cell method 
(see, for example, \cite{allen1989csl}). 
In shock wave simulations the system is not homogeneous: 
near the shock front there is a domain with high density, 
and the common neighbor list algorithm is not efficient for such a situation \cite{rice1998msd}. }
\end{itemize}

Two modifications of the MOLDY code were introduced to fit the
requirements of detonation simulations:

\begin{itemize}
 \item{
MOLDY is constructed to calculate potential energy and forces from a wide set of common analytic potentials. 
We added an option to calculate the potentials and forces from a pre-prepared stored table, 
using cubic spline interpolation.  }
 \item{ 
The initial velocities in MOLDY's simulations are sampled from the Maxwell Boltzmann (MB) distribution. 
The pellet initial velocity were modified so that they could be preassigned,
while the velocities of the slab atoms are sampled from the MB distribution. }
\end{itemize}

\section{The functional form of the exothermic potential}
\label{exo_poten_func_form}
The interatomic potential described in figure \ref{fig:exo_poten}, 
can be constructed by a piecewise defined function,
which is composed from three segments, each having a different functional form:

\begin{widetext}
\begin{equation}
{
V(r) = \left\{ \begin{array}{l@{\ \ \ \ \ \ \ }r}
 D_1 \left( {e^{ - \beta_1\left( {r - r_{min} } \right)}  - 1} \right)^2  + Q & 0 < r < r_{min}  \\ 
 c_3 r^3  + c_2 r^2  + c_1 r + c_0 & r_{min}  < r < r_{bar}  \\ 
 D_2 \left( {1 - \left( {e^{ - \beta_2\left( {r - r_{bar} } \right)} - 1} \right)^2 } \right) & r_{bar}  < r < r_{cut}  \\ 
 \end{array} \right.
}
\label{eq:exo_fuctional_form}
\end{equation}
\end{widetext}

With $r_{min}$ = 1.5 \AA , $r_{bar}$ = 1.9 \AA\ and $r_{cut}$ = 7 \AA\ 
refer to the positions of the local minimum, the barrier, and the cutoff, respectively.
$D_1$ = 0.25 eV is the energy barrier height, 
$Q$ is the energy release during decomposition of the molecule (the exothermicity of the reaction), 
and $D_2$ should satisfy the requirement: $ D_2 = D_1 + Q $, so the function will be continuous.
On the first segment there is a shifted Morse potential, so under small displacements, 
the molecule's behavior is similar to the behavior under standard Morse potential.
The third segment is an inverted Morse potential, and it serves as the repulsive potential at the decomposed state.
The role of the polynomial in the intermediate segment is to link between the two edge segments. 
Therefore, the coefficients $c_0$, $c_1$, $c_2$ and $c_3$, are chosen to make the function and its first derivative continuous.

We determined the polynomial coefficients by the requirements for continuity 
of the potential and its first derivative at the points $r_{min}$ and $r_{bar}$ 
(the derivative vanishes at these points). 
There are four requirements, so the coefficients are determined uniquely. 
Using $Q$ =  1.5 eV and $D_1$ = 0.25 eV (and, consequently, $D_2$ = 1.75 eV), 
and using \AA\ as the length unit, we got
$c_3$ = -7.8125, $c_2$ = 39.8437, $c_1$ =  -66.7969, and $c_0$ =  38.4141, 
with the appropriate units.
One can see that we can obtain continuity of the second derivative as well, 
by choosing $ \beta_1=\sqrt{3}/(r_{bar}-r_{min}) $ = 4.3301 $\text{\AA}^{-1}$ 
and $ \beta_2=\beta_1 \sqrt{D_1/D_2} $ = 1.6366 $\text{\AA}^{-1}$. 
However, this is not essential, since the actual calculations during the simulations were done using cubic spline interpolation. 
This interpolation guarantees continuity of the second derivative.

For other exothermicity values, (i.e. different $Q$), we used the same functional form on the first and second segments, with the same parameters' values, except $Q$ and $c_0$, which determine the exothermicity. On the third segment, we determined the potential values in a way that maid only little gap of the function at $r_{cut}$.
We also constructed potentials with different values for the barrier energy, using the same method.

\section{Crystal Characterization}
\label{sec:cryst_charac}

Before shock propagations are examined the equilibrium structure and
characteristics has to be determined. 
This was carried out by evaluating the position correlation function at different temperatures. 
In addition, the acoustical sound velocity was calculated.

\subsection{Stability and Phase Transition Temperature}
Equilibrium NPT molecular dynamics simulations were performed under constant pressure of 1 bar, 
and at different temperatures. 
Under these conditions the stability of the N-C molecules as well as the stability of the crystal 
and its melting point were evaluated.

We found that the molecules are stable in the temperature range between 0$^\circ$K and 250$^\circ$K. 
When the temperature was raised beyond 250$^\circ$K, some of the N-C bonds started to dissociate. 
This is consistent with the low dissociation barrier of the molecule under examination, which is 0.25 eV.

\begin{figure*}[htbp]  
\begin{center}
$ \begin{array} {cc}
\includegraphics*[width=3.0in]{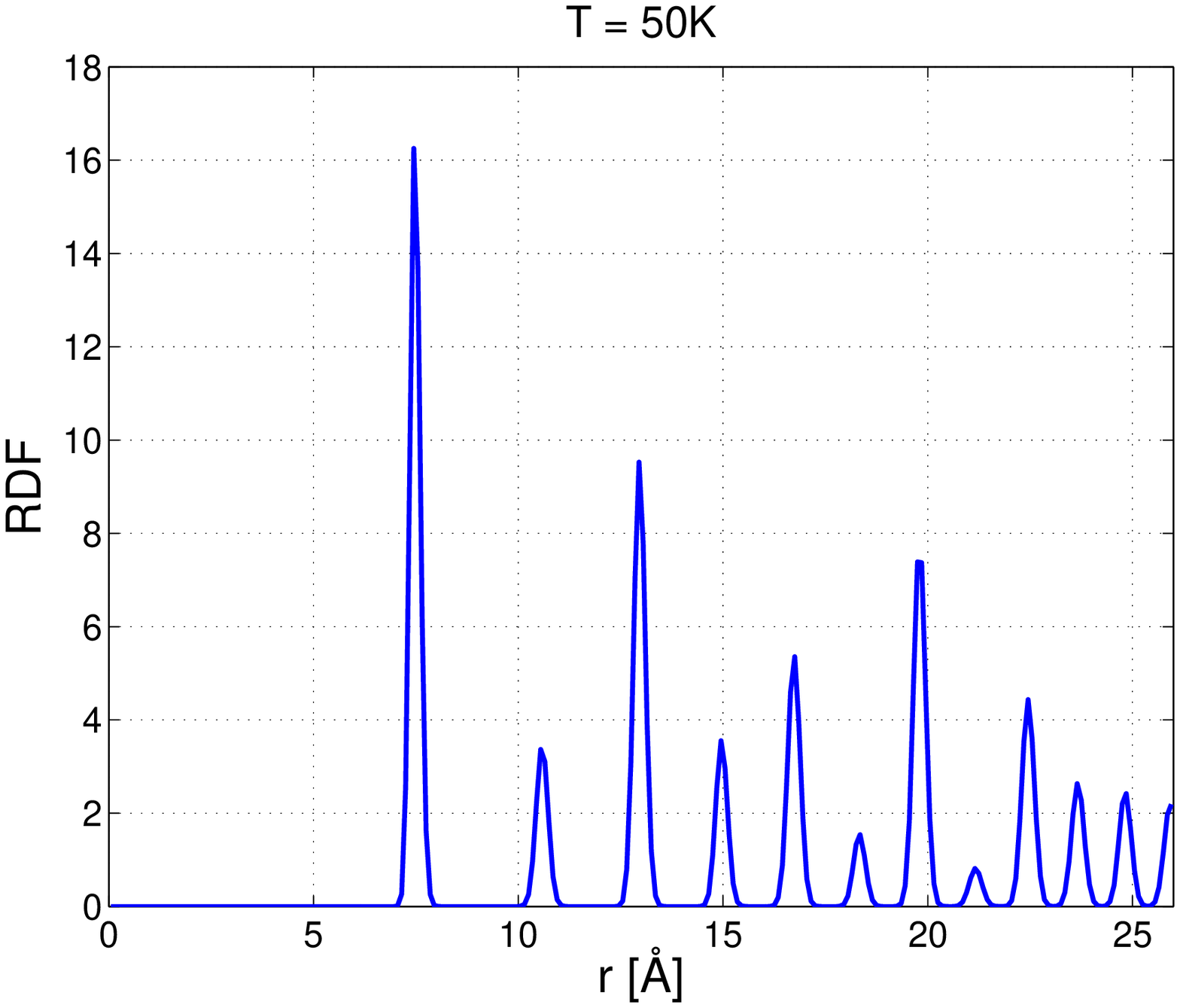} &
\includegraphics*[width=3.0in]{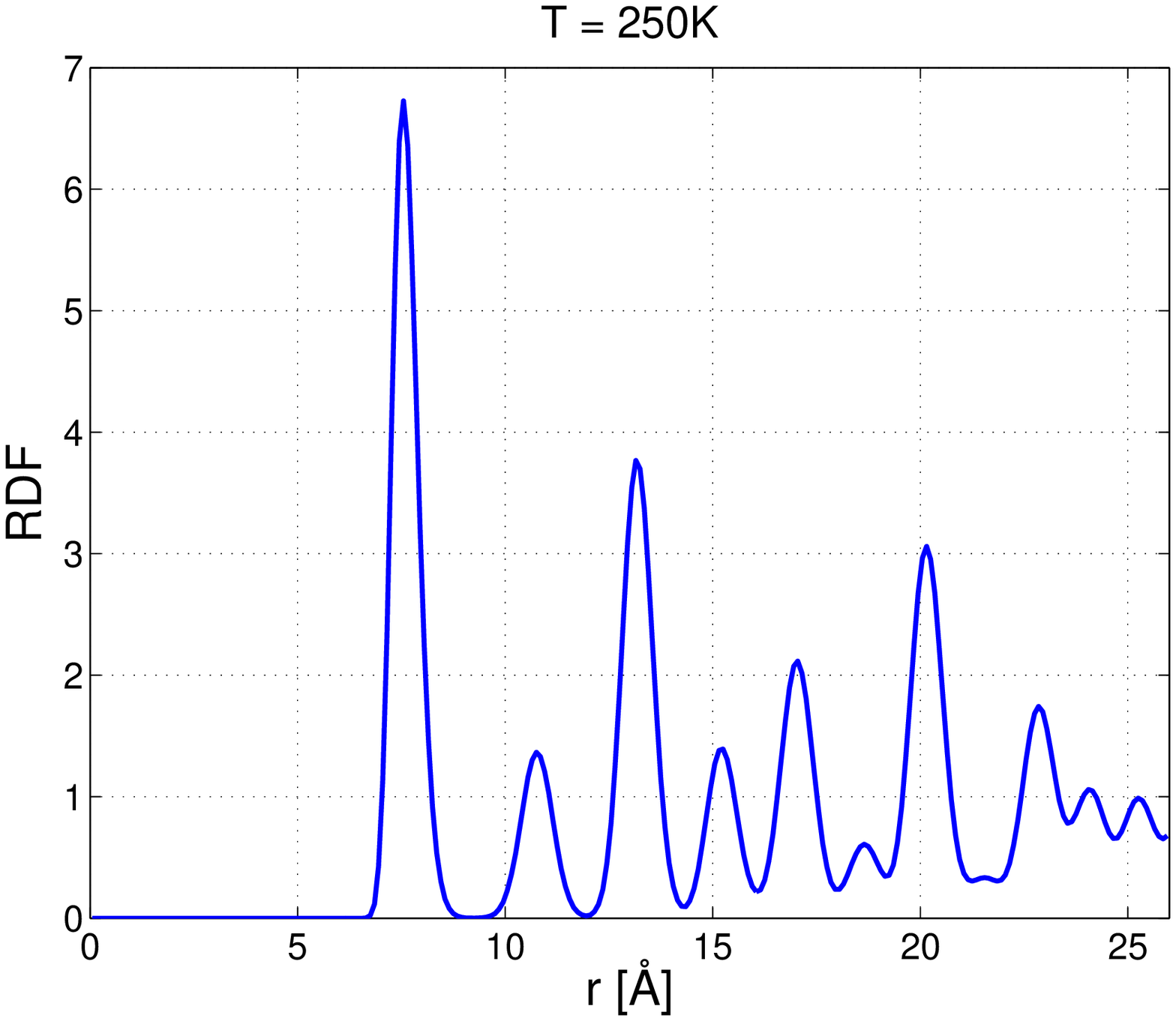} \\
\end{array} $
\end{center}
 \caption{Radial distribution function of the crystal. Results are of a NPT simulations with 504 molecules 
at pressure of 1 bar. The left plot corresponds to 50$^\circ$K, and the right corresponds to 250$^\circ$K.}
 \label{fig:RDF_50_250}
\end{figure*}

The radial distribution functions (RDF) of the N-N pairs were calculated. 
The results of the calculations for 50$^\circ$K and 250$^\circ$K are shown in figure \ref{fig:RDF_50_250}. 
The observed peaks are in positions reflecting the FCC lattice. 
The RDF verify that the crystal is solid up to 250$^\circ$K.
As the temperature is raised  from 50$^\circ$K to 250$^\circ$K, The peaks of the RDF get broadened, 
and at 250$^\circ$K the RDF does not vanish after the third peak, as a result of large fluctuations.

\begin{figure}[htbp]  
\begin{center}
\includegraphics*[width=3.0in]{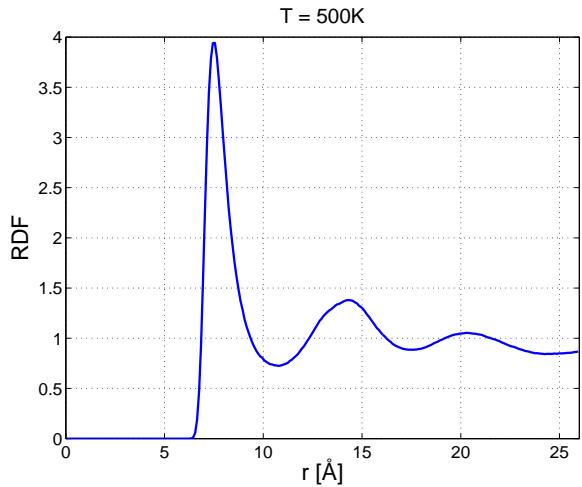} 
\end{center}
 \caption{Radial distribution function  of a NPT simulations with 504 molecules at pressure of 1 bar and 
temperature of 500$^\circ$K.}
 \label{fig:RDF_500}
\end{figure}

On higher temperature, at 500$^\circ$K, some of the molecules decompose and the slab melts. 
The RDF of the system at this temperature shows liquid behavior. 
In figure \ref{fig:RDF_500} there is a plot of the RDF of N-N pairs on 500$^\circ$K. 
There are no sharp peaks at distances greater than the distance of the first nearest neighbors peak. 
This character of the RDF is typical for the liquid phase.

\subsection{Elastic Sound Velocity}
\subsubsection{The elastic sound velocity of our crystal}

The MD simulation was used to determine velocity of the pressure P-wave in the crystal.
A small longitudinal displacement of few crystal layers along the [111] direction was 
used to induce these waves. 
The perturbation propagation velocities was then determined. 
In order to ensure that the displacements are small enough, 
the simulations were repeated with various displacement amplitudes. 
It was then verified that the propagation velocity is independent of the displacement amplitude. 
We found that the P-wave velocity is approximately 2.6 unit cells per picosecond 
($\approx$4.6\e{3} m/s).

\subsubsection{P-waves Velocity on Morse Crystals: A Discussion}
\label{sec:pwave_vel}
A simple 1-D model may give an insight regarding the P-wave velocity in  crystals 
for which the pair interaction is described by the Morse function. 
In this model there is an infinitely long chain of masses. 
Let us consider first an infinite harmonic chain. 
In this case the interaction between neighboring masses is harmonic, 
with a spring constant $\alpha$. 
We denote the distance between two neighboring masses $l$. 
The equation of motion for the n'th mass is:
\begin{equation}
m\ddot x_n  = \alpha (x_{n + 1}  - 2x_n  + x_{n - 1} ) 
\end{equation}
If we assume that the time dependent position is given by:
\begin{equation}
x_n (t) = (A_ +  e^{ + iknl}  + A_ -  e^{ - iknl} )\cos (\omega t + \phi ) 
\end{equation}
we get the dispersion relation:
\begin{equation}
\omega (k) = 2\omega _0 \sin \left( {\frac{{kl}}{2}} \right) 
\end{equation}
where we defined  $ \omega _0  = \sqrt {\alpha / m} $.
The p-wave velocity for this case is 
\begin{equation}
v_s  = \left. {\frac{{d\omega }}{{dk}}} \right|_{k = 0}  = \omega _0 l\cos \left. {\left( {\frac{{kl}}{2}} \right)} \right|_{k = 0}  = \omega _0 l = \sqrt {\frac{\alpha }{m}} l
\end{equation}

This model can be used to describe the harmonic approximation of small
vibrations around the equilibrium distance in the Morse potential (and
other potentials). If the vibration amplitude is small
enough for the harmonic approximation to hold, the propagation velocity
of small perturbations should be independent of the perturbation
amplitude. In such cases, we substitute $ \alpha  = 2\beta^2 D $, 
the second derivative near the minimum of the potential. 
Now the p-wave velocity is 
\begin{equation}
v_s =  \sqrt {\frac{\alpha }{m}} l = \sqrt {\frac{2 D}{m}} \beta l
\end{equation}

This result suggests that the p-wave velocity scales linearly with $\beta$.

\begin{figure}[htbp]  
 \centering \includegraphics*[width=3.0in]{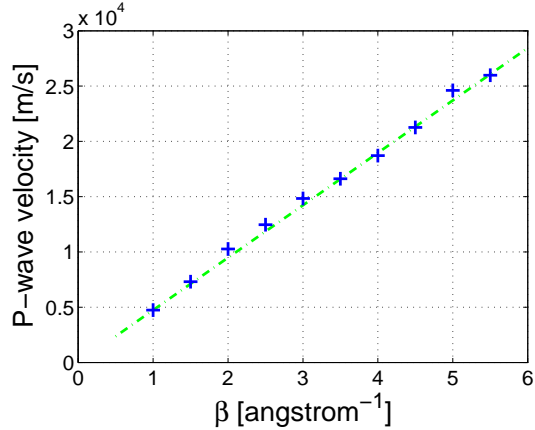}
 \caption{P-wave velocity in the crystal as a function of the $\beta$ 
 parameter in the intermolecular Morse potential.
Each velocity was calculated by simulating propagation of small displacements in a crystal,
with the relevant $\beta$ parameter,
 and marked in the figure by a cross.
The dashed line is an extrapolation to small $\beta$ values assuming linear scaling.}
 \label{fig:p_bet}
\end{figure}

The P-wave velocity in this simple 1-D model is not directly comparable 
to the P-wave velocity in the 3-D fcc slab, 
since the elastic waves on anisotropic crystal is determined by 
the elastic tensor $\sigma$, which has several parameters \cite{carcione}.
However, some insights can be gained:
The MD simulation was employed to evaluate the P-wave velocity along 
the [111] direction of our slab.
The assumption of the velocity's linear scaling as a function of $\beta$ 
is tested by simulating propagation of small displacements in different crystals, while changing $\beta$.
The results are shown in figure \ref{fig:p_bet}, marked by crosses. 
The dashed line in this figure is an extrapolation of the $\beta$ = 1 case, assuming linear scaling. 
We can see that our assumption is valid in a wide range of the $\beta$ values.

\section{The transition to stationary detonation waves}

\label{sec:impact_indepen}

\begin{figure}  
 \centering \includegraphics*[width=3.0in]{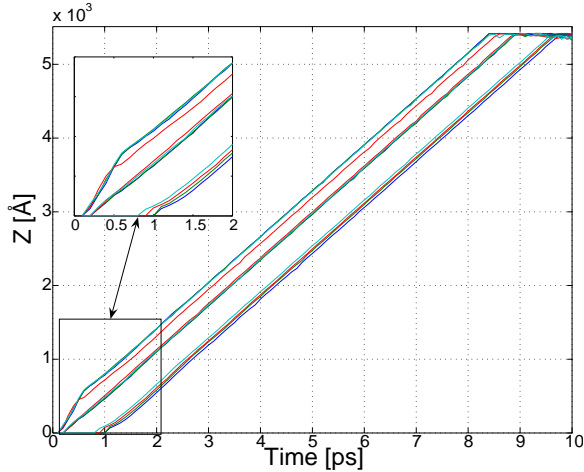}
  \caption{
Reaction front location versus time in simulations with three different 
initial velocities of the pellet. The location was determined by a significant dissociation ratio.
The results are  grouped in three categories: 
The rightmost bundle represents five different simulations with the lowest pellet
velocity used ($\approx$10 km/s). 
The central bundle corresponds to simulations (5 runs)
with medium pellet velocity ($\approx$50 km/s), 
while the leftmost bundle corresponds to
simulations with the largest pellet velocity ($\approx$120 km/s). 
(These high values of initial velocities are used for demonstration: 
pellet velocities of 10, 50, and 120 km/s give rise to initial transient shock waves 
that are, respectively, slower, similar, and higher than the developing stable detonation wave, 
which is approximately 65 km/s.
Smaller values of pellet's velocities were used throughout most of this paper).
The different simulations in each one of the
three groups correspond to different initial conditions used (slightly
different initial pellet-crystal distance). 
A short time after the pellet collision with the slab, 
different shock wave velocities develop in each simulation.
The differences between the
trajectories at short times are shown in the insert of the figure.
A stable, steady state, velocity is reached after an additional propagation period. 
The shock velocity depends only on crystal parameters, 
which are identical in all simulations.
}
 \label{fig:diff_init}
\end{figure}

A stable detonation wave is independent of initial conditions.
To check this hypothesis different initial pellet velocities were tested.
The location of the
reaction front was defined as the first point where molecular
decomposition is identified. 
Figure \ref{fig:diff_init} shows the location of the reaction front
as a function of time in simulations with three different initial pellet velocities. 
Inspection of these results shows that a short time after the pellet
collision with the slab, different shock wave velocities develop in
each simulation. 
This is a transient: a stable, steady state, velocity is reached
after a short additional propagation period. 
The steady state shock velocity depends only on crystal parameters. 
Therefore, it is clear that the pressure wave induced by the pellet in this model crystal 
develops into a constant velocity detonation wave. 
A similar transition from initiation dependent velocity to a steady state
velocity that depends only on crystal parameters was previously reported
\cite{PhysRevB.33.2350}.

\bibliography{WeakDetRef} 
\bibliographystyle{unsrt}

\end{document}